\documentclass[twocolumn,superscriptaddress]{revtex4-1}

\usepackage[a4paper,left=2.0cm,right=1.9cm,top=2.2cm,bottom=2.2cm]{geometry}
\usepackage{graphicx}
\usepackage{amsmath}
\usepackage{amssymb}
\usepackage{color}
\usepackage{bbm}
\usepackage[normalem]{ulem}
\usepackage{paralist,hyperref} 
\usepackage{lipsum}

\usepackage{titlesec}
\titlespacing{\section}{0ex}{5ex}{1.5ex}
\titlespacing{\subsection}{0ex}{3ex}{1ex}
\titleformat{name=\section}{\centering \bfseries \large}{\thesection.}{.5em}{}
\titleformat{name=\subsection}{\bfseries \normalsize}{\thesection~\thesubsection.}{.2em}{}

\def\App{{Appendix}}
\def\d{\!\!\!\mathrm{d}}
\def\tr{\mathrm{Tr}}
\def\D{{\cal D}}

\def\stst{{\varrho}^{ss}} 
\DeclareMathOperator{\sech}{sech}
\def\sigmar{\sigma_{\hat{\mathbf{r}}}}
\def\lowT{\lim_{\mbox{\it \footnotesize low\,}T} \rho^{(2)}_S}

\newcommand{\Gib}{{Gibbs}}
\newcommand{\MFstate}{MFG~state}

\begin{document}

\title{Weak and ultrastrong coupling limits of the quantum mean force \Gib~state}

\author{J. D. Cresser}
\email[]{j.d.cresser@exeter.ac.uk}
   \affiliation{Department of Physics and Astronomy, University of Exeter, Stocker Road, Exeter EX4 4QL, UK.}
   \affiliation{School of Physics and Astronomy, University of Glasgow, Glasgow, G12 8QQ, UK.}
   \affiliation{Department of Physics and Astronomy, Macquarie University, 2109 NSW, Australia.}

\author{J. Anders}
\email[]{janet@qipc.org}
   \affiliation{Department of Physics and Astronomy, University of Exeter, Stocker Road, Exeter EX4 4QL, UK.}
    \affiliation{Institut f\"ur Physik und Astronomie, University of Potsdam, 14476 Potsdam, Germany.}

\date{\today}

\begin{abstract}

The Gibbs state is widely taken to be the equilibrium state of a system in contact with an environment at temperature $T$. 
However, non-negligible interactions between system and environment can give rise to an altered  state. 
Here we derive general expressions for this mean force Gibbs state, valid for any system that interacts with a bosonic reservoir. 
First, we derive the state in the weak coupling limit and find that, in general, it maintains coherences with respect to the bare system Hamiltonian. 
Second, we develop a new expansion method suited to investigate the ultrastrong coupling regime. This allows us to derive the explicit form for the mean force Gibbs state, and we find that it becomes diagonal in the basis set by the system-reservoir interaction instead of the system Hamiltonian.
Several examples are discussed including a single qubit, a three-level V-system and two coupled qubits all interacting with bosonic reservoirs. 
The results shed light on the presence of coherences in the strong coupling regime, and provide key tools for nanoscale thermodynamics investigations. 

\end{abstract}

\maketitle

The last decade has seen much progress in building a comprehensive framework of ``strong coupling thermodynamics" \cite{Miller2018} that extends standard thermodynamic relations to  take into account the impact of system-environment interactions \cite{Jarzynski2004a,Campisi2009a,Campisi2009b,Gelin2009,Hilt2011,Hilt2011a,Seifert2016,Philbin2016,Jarzynski2017,Aurell2017,Strasberg2017b, Miller2017,Aurell2018,Schaller2018,Strasberg2018b,Miller2018a,Strasberg2019,Correa2017,Hovhannisyan2018a,Perarnau-Llobet2018,Huang2020,Rivas2020}. 
Based on the formal concept of the Mean Force Gibbs (MFG) state \cite{Kirkwood1935,Miller2018}, 
strong coupling thermodynamic potentials have been identified \cite{Seifert2016,Philbin2016,Jarzynski2017,Aurell2017,Aurell2018}, 
detailed entropy fluctuation relations have been shown to hold \cite{Strasberg2017b, Miller2017},
and quantum measurements have been included  in a stochastic description of strongly coupled quantum systems \cite{Strasberg2019}. 
In quantum thermometry, strong coupling has been found to  improve measurement precision \cite{Correa2017,Hovhannisyan2018a}, while it can be detrimental for the efficiency of quantum engines \cite{Perarnau-Llobet2018}.

For classical nanoscale systems, the impact of the environment, or reservoir, beyond setting the system's temperature has been known since the 30's~\cite{Kirkwood1935}. System-reservoir interactions lead to a modification of the system's bare Hamiltonian, $H_S$, to an effective (mean force) Hamiltonian which is routinely calculated for classical systems in chemistry simulations \cite{Roux1995}. The resulting classical distribution or quantum state of the system, known as the  \MFstate, is the reduced state
\begin{equation} \label{eq:DefMFGS}
	\rho_S=\tr_R\left[\tau_\text{SR}(\beta)\right]=\tr_R\left[\frac{e^{-\beta H_\text{SR}}}{Z}\right]
\end{equation}
of the global  system-plus-reservoir \Gib~state $\tau_{SR}(\beta)$ at inverse temperature $\beta =1/k_B T$. The predictions of strong coupling thermodynamics differ from standard thermodynamics because the \MFstate~can deviate significantly from the standard \Gib~state $\tau_S (\beta) \propto e^{-\beta H_S}$ widely used across all of the natural sciences. Apart from leading to corrections to the state's probabilities, the system-reservoir coupling can lead to $\rho_S$ maintaining coherences with respect to the basis of $H_S$ at low and intermediate temperatures \cite{Purkayastha2020}.
This is significant because coherences are often viewed as indication of the quantumness of a system and considered a quantum `resource' \cite{Streltsov2017}. Beyond quantum thermodynamics \cite{Uzdin2015,Kammerlander2016,Francica2020,Messinger2020,Purkayastha2020,Hammam2021}, coherences play an important role in some biological processes \cite{Lloyd2011,Lambert2013,Jeske2015,Dodin2016,Dodin2018}, and may also affect a material's magnetization behaviour  \cite{Anders2020}.

But beyond a few limited examples, the explicit evaluation of the reduced state $\rho_S$ has generally proven intractable. This lack of immediately applicable expressions of $\rho_S$ severely hampers progress in applying strong coupling thermodynamics methods to concrete systems, as well as characterising thermodynamic properties of strongly coupled equilibrium states, including the presence of coherences.

Here we provide explicit forms of the \MFstate~for \emph{general quantum systems} coupled to bosonic environments, in the weak and ultrastrong coupling limit, respectively.  For the previously unexplored ultrastrong coupling regime, we develop a new perturbative approach which leads to a surprisingly neat expression for the MFG state. For the weak coupling limit we give concrete conditions on the coupling strength $\lambda$ for the coupling to be considered ``weak''.
Several representative examples are discussed in both regimes, including a single qubit, a three-level V-system and two coupled qubits all interacting with bosonic reservoirs.

\smallskip

\emph{General setting.~}
Throughout we consider global equilibrium states $\tau_\text{SR} (\beta) \propto e^{-\beta H_\text{SR}}$ of a system $S$  coupled to bosonic reservoir(s) $R$. For a single continuous reservoir, the full Hamiltonian $H_{SR}$ is~\cite{Newman2017,Nazir2018}
\begin{equation}\label{eq:totalHamiltonian}
	H_{SR}=H_S+ \int_{0}^\infty \d\omega \, \frac{p^2(\omega)+\left(\omega q(\omega)+\lambda {\sqrt{\frac{2J(\omega)}{\omega}}} X \right)^2}{2} ,
\end{equation}
where $\left[q(\omega),p(\omega')\right]=i \, \delta(\omega-\omega')$ are the commutation relations for reservoir position and momentum operators~\footnote{For $H_S$ bounded from below, $H_{SR}$ will clearly also be bounded from below. However, if one took instead the form $H_{SR}=H_S+H_R+\lambda X B$ \cite{Ford1997,Nazir2018} then energies unbounded from below can arise at strong coupling for systems of infinite dimension. Care must be taken when applying the results derived here in such situations.}. We choose units with $\hbar =1$ and $k_B=1$ throughout. The strength of the system-reservoir coupling is scaled by a dimensionless factor $\lambda$, where the coupling is to an arbitrary dimensionless system operator $X=X^\dagger$, and $J(\omega)$ is a real function of $\omega$ 
which will later be identified as the reservoir spectral density. An extension of \eqref{eq:totalHamiltonian} to two reservoirs will also be considered.

While for $\lambda \to 0$  the \MFstate~$\rho_S$ clearly reduces to $\tau_S(\beta)= {e^{-\beta H_S} /Z_S}$ with $Z_S = \tr \left[e^{-\beta H_S} \right]$,  it will differ considerably from $\tau_S$ for non-negligible $\lambda$.  Here we solve this problem for a number of physically meaningful cases.
To prepare the derivation, we expand $H_{SR}=H_S'+H_R+\lambda V$ as a sum of a reservoir Hamiltonian $H_R=\int_{0}^\infty \d \omega \, \omega \, \, {\left(b^\dagger(\omega) b(\omega)+b(\omega)b^\dagger(\omega)\right) / 2}$ with $b(\omega)=\sqrt{\omega/2}\left(q(\omega)+ip(\omega)/\omega\right)$ reservoir annihilation operators, an interaction energy $\lambda \, V = \lambda \, X B$ with $B=\int_{0}^{\infty} \d \omega \, \sqrt{J(\omega)}\left(b(\omega)+b^\dagger(\omega)\right)$,  and an effective system Hamiltonian $H_S'=H_S+\lambda^2X^2Q$. Here $Q=\int_{0}^{\infty}d\omega J(\omega)/\omega $ is the reorganization energy \cite{Wu2010a,Ritschel2011,Fruchtman2016}. The latter has a non-trivial impact on $\rho_S$ only if $X^2 \not\propto\mathbb{I}$. I.e., for many qubit problems, for which $X \sim \sigmar$ for some Pauli-matrix $\sigmar$, the reorganization energy can be disregarded as a constant off-set. 
Before discussing the different coupling limits, we first comment on the high temperature limit at all finite coupling strengths.

\smallskip
 
\emph{High temperature limit.~} For $\beta\to 0$ the trace over the reservoir in $\rho_S =\tr_R[\tau_{SR} (\beta)]$ can be  performed directly using a factorization approximation of $\exp[-\beta H_{SR}]$, see \App~\ref{app:factorization}. One finds the cancellation of the reorganization energy term in $H'_S$,  yielding $\rho_S =\tau_S (\beta) + {\cal O}(\beta^2)$, i.e. the system's Gibbs state with respect to the bare Hamiltonian $H_S$ emerges. 

\smallskip

\emph{Weak coupling.~} 
We now turn to arbitrary temperatures and consider the weak coupling limit, quantitatively defined  in Eq.~\eqref{eq:binomialApproxConditionmain} below. To obtain $\rho_S$, we write the system operator $X$ in terms of the energy eigenoperators $X_n$ for the system, i.e., $X=\sum_{n}X_n$ where $n$ ranges over positive and negative values. The $X_n$  are defined through $\left[H_S,X_n\right]=\omega_nX_n$ with $\omega_n$ the Bohr frequencies. Since $X=X^\dagger$ one has $X_n=X_{-n}^\dagger$ and $\omega_n=-\omega_{-n}$, with $\omega_0=0$. Using the Kubo expansion we obtain the  \MFstate,  correct to second order in coupling $\lambda$ \cite{Mori2008, Fleming2011,Thingna2012,Subasi2012,Purkayastha2020} indicated by the superscript~${}^{(2)}$,
\begin{eqnarray}\label{eq:NormalizedMFGSmain}
	\rho_S^{(2)}
	=\tau_S 
	&&+\lambda^2\beta\sum_{n}\tau_S\left(X_nX_n^\dagger
	- \tr_S \left[\tau_SX_nX_n^\dagger\right]\right) \D_{\beta}(\omega_n) \nonumber \\
	&&+\lambda^2\sum_{n}
	\left[X_n^\dagger, \tau_SX_n\right]
	\frac{d\D_{\beta}(\omega_n)}{d\omega_n} \\
	&&+\lambda^2\sum_{m\ne n}\left(
\left[X_m,X_n^\dagger \tau_S\right] +\text{H.c.}\right) \frac{\D_{\beta}(\omega_n)}{\omega_{mn}} , \nonumber 
\end{eqnarray}
where $\omega_{mn} = \omega_m - \omega_n$ are frequency differences, the double sum is over all ordered pairs $(m, n)$ with $m\neq~n$, and derivation details are given in \App~\ref{app:ArbTempWeakCoup}. The temperature dependent coefficient $\D_{\beta}(\omega_n)$ includes generally principal part integral transforms of reservoir correlation functions involving $J(\omega)$ and Bose number statistics $n_{\beta}(\omega)$. 
These integrals are responsible for population-coherence coupling, terms that are routinely ignored in the Bloch-Redfield master equation description of the dynamics, in which case the steady state simplifies to $\tau_S$ \cite{Agarwal2001}. The impact of these terms here is that the reduced state \eqref{eq:NormalizedMFGSmain} can differ very significantly from  $\tau_S$. 
In particular, $\rho^{(2)}_S$ may maintain energetic coherences (coherences in the basis of the bare Hamiltonian $H_S$) since the commutator  
\begin{equation} \label{eq:weakcouplingcoherences}
	[\rho^{(2)}_S, H_S] 
	= \lambda^2\sum_{m\ne n} 
	\left( \left[X_m,X_n^\dagger \tau_S\right] +\text{H.c.} \right) \, \D_{\beta}(\omega_n), \quad
\end{equation}
is in general non-trivial. 

We highlight that the derivation of \eqref{eq:NormalizedMFGSmain} requires 
\begin{equation}\label{eq:binomialApproxConditionmain}
	|\lambda| \ll \frac{1}{\sqrt{| \beta\sum_{n}\tr_S\left[\tau_SX_nX_n^\dagger\right] \D_{\beta}(\omega_n)  | }}
\end{equation}
to be valid, see \App~\ref{app:ArbTempWeakCoup}. 
Beyond the loose requirement that $\lambda$ ought to be ``small'', this condition gives a well-quantified limit for $\lambda$ being in the weak coupling regime at a given $\beta$. Note that the range of $\lambda$ for which the weak coupling regime and hence \eqref{eq:NormalizedMFGSmain} is applicable changes as a function of temperature, with larger temperature generally allowing larger $\lambda$.

 
\smallskip

As first example for expression \eqref{eq:NormalizedMFGSmain}  we consider the spin-boson model, i.e. a {\bf Single qubit} with Hamiltonian $H_S=\omega_q\sigma_z/2$, coherently coupled to a bosonic bath with $X=\cos\theta \,\, \sigma_z-\sin\theta \,\, \sigma_x$. 
This model describes a charge qubit in a double quantum dot \cite{Purkayastha2020}, as well as the stochastic behaviour of spins in magnetic materials \cite{Anders2020}.
Coherences have recently been identified in the \MFstate~for this example \cite{Purkayastha2020}, and our expression \eqref{eq:NormalizedMFGSmain} reproduces these results, as detailed in \App~\ref{App:spinboson}.

\smallskip

As a novel illustration of the power of \eqref{eq:NormalizedMFGSmain} we consider the three-level {\bf V-system} with Hamiltonian $H_S =  0 \, |0 \rangle \langle 0| + \omega_1 \, |1 \rangle \langle 1|+  \omega_2 \, |2 \rangle \langle 2|$ with $\omega_{1,2} =\omega_q \mp \Delta/2$ and $\Delta/2 \ll\omega_q$, coupled to the reservoir via $X = \sqrt{2} \, |\psi\rangle\langle 0|+ h.c.$, where $|\psi\rangle = (|1\rangle+|2\rangle)/\sqrt{2}$. These systems can represent biomolecules and have attracted significant attention as their dynamics, according to a Bloch-Redfield master equation, gives rise to metastable noise-induced energetic coherences~\cite{Tscherbul2014}. 
These dynamical coherences are long lived, but they eventually decay. However, the Bloch-Redfield approach makes approximations which, despite non-negligible environment coupling, force $\tau_S$ to be the steady state from the outset.  In contrast, when the reservoir impact is included in the  form of the MFG state $\rho^{(2)}_S$, energetic coherences persist in the V-system even in equilibrium.

\begin{figure}[t]
\includegraphics[trim=1.15cm 0.1cm 0cm 0cm,clip, width=0.48\textwidth]{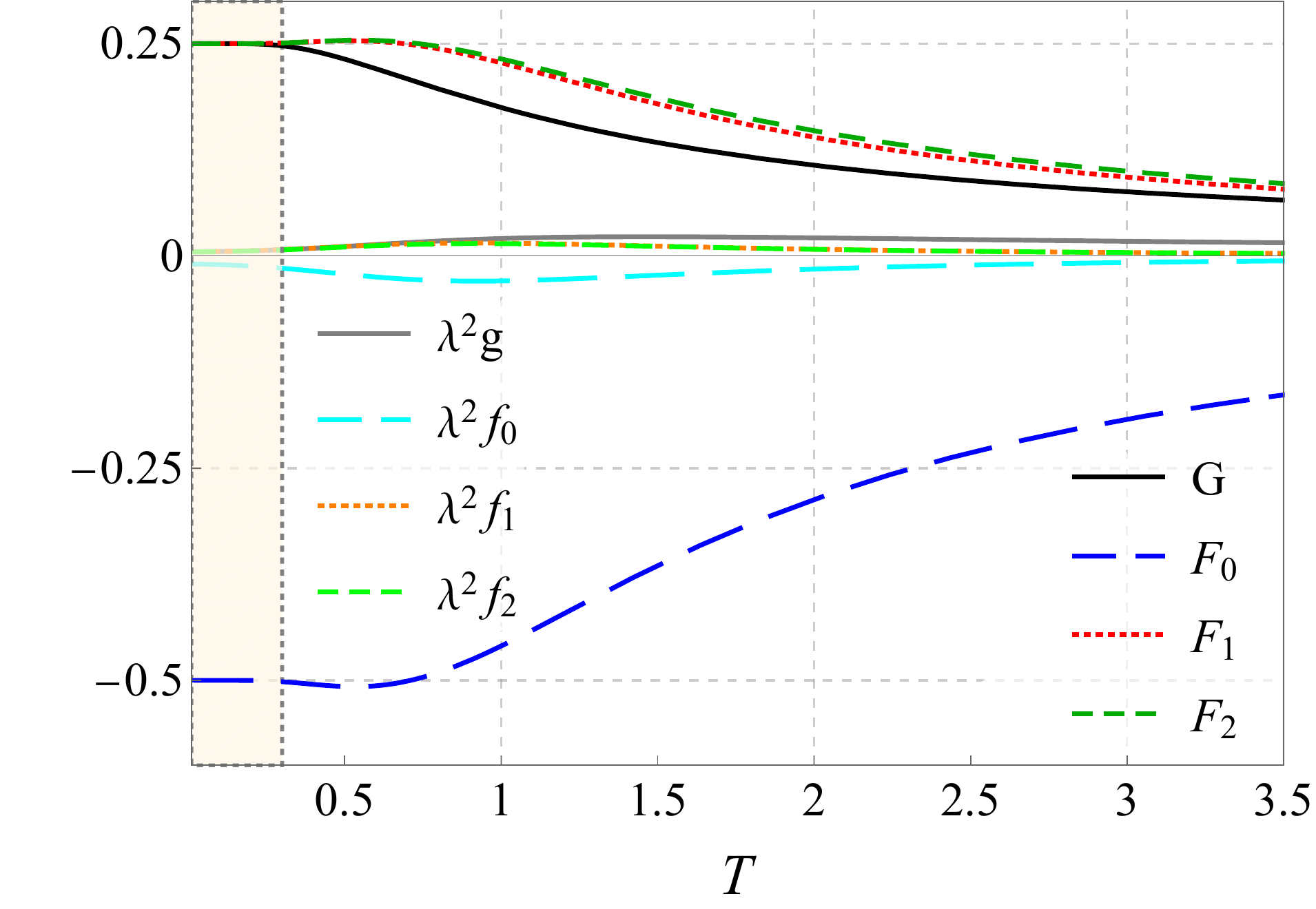}
\caption{\label{fig:Vcoh} Mean force \Gib~state corrections for the V-system as a function of temperature $T$.  In the weak coupling limit, with $\lambda = 0.1$, the $|1\rangle\langle 2|$-coherence is $\lambda^2 g$ (solid grey) and the population corrections are $\lambda^2 f_p$ for the three energy eigenstates $p=0,1,2$ (cyan large-dashed, orange dotted, light green dashed), see Eq.~\eqref{eq:Vsystem}. The weak coupling result is valid in the white-shaded temperature range, where the low temperature condition \eqref{eq:binomialApproxConditionmain} is obeyed.
In the ultrastrong coupling limit, $\lim_{\lambda \to \infty}$, the \MFstate~$\rho_S$ is given by \eqref{eq:Vstrong}, a specific case of \eqref{eq:PartitionedState}. Its coherence is $\mbox{\it G} = \lim_{\lambda \to \infty}\langle 1|\rho_S|2\rangle+h.c.$ (solid black) and the population corrections are $F_p = \lim_{\lambda \to \infty}\langle p|\rho_S|p\rangle- \langle p|\tau_S|p\rangle$   (blue large-dashed, red dotted, green dashed), evidencing significant deviations of $\rho_S$ from $\tau_S$.
Parameters for this plot:  $\omega_q=3$ and $\Delta =0.1$, and the coupling strength to the bosonic environment in \eqref{eq:totalHamiltonian} is set by the spectral density $J(\omega) = Q\,\tau_c  \, \omega \, e^{-\tau_c \, \omega}$ with $Q = 10$ and $\tau_c = 1$.}
\end{figure}

To obtain $\rho^{(2)}_S$ we identify the eigenoperators and Bohr frequencies $(X_n,\omega_n)$ as $(|1\rangle\langle 0|, \omega_1)$, $(|2\rangle\langle 0|, \omega_2)$, $(|0\rangle\langle 2|,-\omega_2)$ and $(|0\rangle\langle 1|, -\omega_1)$. Substituted into \eqref{eq:NormalizedMFGSmain} one obtains
\begin{align}\label{eq:Vsystem}
	\rho_S^{(2)}
	= \tau_S 
	&+ \lambda^2 \sum_{p=0,1,2} \, f_p(\beta) \, |p \rangle  \langle p |   \nonumber\\
	&+ \lambda^2 \, g(\beta) \,  \left(|1\rangle\langle 2|+|2\rangle\langle 1|\right). 
\end{align}
Expressions for $\lambda^2g(\beta)$, as well as the diagonal coupling corrections $\lambda^2f_p(\beta)$ are given in Eq.~\eqref{eq:gandf},  and plotted as a function of temperature $T$ in Fig.~\ref{fig:Vcoh}. Of particular significance is the presence of non-vanishing coherence $g(\beta)$ between the upper levels $|1\rangle$ and $|2\rangle$. At low temperatures coherence arises due to the environment's vacuum fluctuations, while it depletes at higher temperature due to classical fluctuations. Coherence in fact peaks at an intermediate temperature whose scale is set by $\omega_q$, as confirmed numerically.

\smallskip

\emph{Ultrastrong coupling.~} 
To derive the \MFstate~$\rho_S$ in the ultrastrong coupling limit $\lambda \to \infty$, perturbative techniques following the weak system-reservoir coupling are inapplicable and an entirely new approach is required. 
We here develop such a new approach, see   \App~\ref{app:StrongCouplingGibbsStateDetails}, by reversing the roles of system and perturbation via $H_S'+H_R+\lambda V\to\lambda(V+\lambda^{-1}(H_S'+H_R))$. 
Building on techniques previously used to study unitary Zeno dynamics \cite{Burgarth2018,Misra1977,Facchi2003,Facchi2003a}, we here apply it in a very different context: in the temperature domain and generalised to open systems, which requires carrying out a highly non-trivial trace over the reservoir.
Importantly, we show that all diverging terms with positive powers of $\lambda$ cancel. 
Equipped with our ultrastrong expansion method one finds, for any quantum system coupled to a bosonic reservoir, the \MFstate
\begin{equation}\label{eq:PartitionedState}
	\lim_{\lambda \to \infty} \rho_S=\frac{e^{-\beta\sum_{n}P_n H_S P_n} }{\tr\left[e^{-\beta\sum_{n}P_n H_S P_n}\right]},
\end{equation}
where $P_n=|x_n\rangle\langle x_n|$ are projection operators on the non-degenerate eigenstates $|x_n\rangle$ of $X$. 
This general analytical form of the \MFstate~in the ultrastrong limit is the main finding of the paper. To our knowledge, it constitutes a completely new result not previously suggested or proven in the literature, not even for specific systems.

The interpretation of the \MFstate~in \eqref{eq:PartitionedState} is that it is still a Gibbs state, but with respect to an effective Hamiltonian  $\sum_{n}P_n H_S P_n$. The impact of ultrastrong coupling is to force the system to equilibrate according to the eigenstates $P_n$ of the now dominant system interaction operator $X$, rather than w.r.t. to the system Hamiltonian $H_S$. Analogously to the standard Gibbs state $\tau_S \propto e^{- \beta H_S} =  \sum_k e^{- \beta E_k} |e_k \rangle \langle e_k| $ where the weights in the exponents are given by the mean value $E_k = \langle e_k| H_S |e_k \rangle$  of the system Hamiltonian in the energy eigenbasis $|e_k\rangle$, the ultrastrong \MFstate~has weights in the exponent that are given by the mean value of the effective Hamiltonian in the effective eigenbasis $|x_n\rangle$, i.e. $\langle x_n| \sum_{m}P_m H_S P_m |x_n \rangle = \langle x_n|  H_S  |x_n \rangle$.
While the derivation of \eqref{eq:PartitionedState} assumes non-degenerate eigenstates $|x_n\rangle$ \footnote{We note that, mathematically, in the ultrastrong limit, the \Gib~state $\tau_S$ is not recovered in the high temperature limit ($\beta \to 0$), due to issues of the order of limits. However, physically a situation where a quantum system couples ultrastrongly to a reservoir at ultrahigh temperatures, is not realistic, so we leave this mathematical issue unresolved here.}, extensions to degenerate eigenspaces are straightforward.

We note that Kawai and coworkers have recently \emph{conjectured} the form of the system's dynamical state at long times \cite{Goyal2019,Orman2020}, i.e., \emph{the  steady state} $\stst_S (t \to \infty)$, to be $\varrho_S^{conj}=\sum_{n}P_n \, \tau_S \, P_n$. This form looks similar to our derived \MFstate~$\rho_S$ in \eqref{eq:PartitionedState} but is structurally different, as exemplified below. 

\smallskip

As an illustration of Eq.~\eqref{eq:PartitionedState} we consider again the {\bf Single qubit}, now ultrastrongly coupled to a reservoir with $X=\boldsymbol{\sigma}\cdot\hat{\mathbf{r}}=\sigmar$ with $\hat{\mathbf{r}}$  an arbitrary unit vector. With the projectors on the eigenstates of $\sigmar$, $P_0 = |+_{\hat{\mathbf{r}}}\rangle\langle+_{\hat{\mathbf{r}}}|$ and $P_1 = |-_{\hat{\mathbf{r}}}\rangle\langle-_{\hat{\mathbf{r}}}|$, the partitioned Hamiltonian becomes $\sum_{n}^{}P_nH_SP_n= \sigmar  \, \cos(\theta) \, \omega_q/2$ where $\hat{\mathbf{r}}\cdot\hat{\mathbf{z}}=\cos(\theta)$.  
A straightforward calculation then gives the \MFstate, see \App~\ref{sub:singlequbit},
\begin{equation} \label{eq:singlequbit}
	\lim_{\lambda \to \infty}\rho_S
	=\tfrac{1}{2}\left(1- \sigmar \, \tanh\left(\tfrac{1}{2} \beta\omega_q \cos(\theta)\right)\right).
\end{equation}
Being diagonal in the basis of the coupling-operator $X$,  for  $\cos(\theta)\neq0$ this state clearly maintains  coherences with respect to the Hamiltonian's $\sigma_z$ basis.

For comparison we note that the conjectured steady state for this system is $\varrho_S^{conj}= \tfrac{1}{2}\left(1-  \sigmar \, \cos(\theta)\tanh\left(\tfrac{1}{2} \beta\omega_q \right)\right)$~\cite{Orman2020}, which differs from \eqref{eq:singlequbit} in the positioning of $\cos (\theta)$. 
Evidence,  based on numerically solving hierarchical equations of motion (HEOM), that the dynamical steady state $\stst_S$ is  numerically close to $\varrho_S^{conj}$ has been provided  \cite{Orman2020}. However, for the specific inverse temperature used in the numerics, the difference between $\varrho_S^{conj}$ and the \MFstate~\eqref{eq:singlequbit} is negligible, and hence the convergence could equally be to \eqref{eq:singlequbit} instead,  see \App~\ref{sub:singlequbit}. 


\smallskip

A second example of the application of Eq.~\eqref{eq:PartitionedState} is  the three-level  {\bf V-System} for which the ultrastrong coupling \MFstate~is given in \App~\eqref{eq:Vstrong}. Its deviations from $\tau_S$ are plotted in Fig.~\ref{fig:Vcoh}, together with the weak coupling corrections given in \eqref{eq:Vsystem}. Coherence between the excited states, $|1\rangle$ and $|2\rangle$, is found to persist at low and intermediate temperatures $T \lesssim \omega_q$. 
As expected, the corrections in the ultrastrong limit are larger in magnitude than those at weak coupling.

\smallskip

The ultrastrong coupling derivation resulting in  Eq.~\eqref{eq:PartitionedState} can  further be extended to situations involving \emph{two reservoirs}, both at the same inverse temperature $\beta$. Here we consider two systems  $S_1$ and $S_2$ (such as qubits), that  interact with each other via $H^{int}_{12}$ as well as each coupling to a bath, through $\lambda_1 B_1X_1$ and $\lambda_2 B_2X_2$, respectively. This gives the total Hamiltonian
$H_{\lambda_1\lambda_2}=H'_S +\lambda_1 B_1X_1+\lambda_2 B_2X_2 + H_B$ with two-system Hamiltonian  $H'_S =H_{1}+H_{2}+H^{int}_{12} +\lambda_1^2 X_{1}^2 Q_1+\lambda_2^2 X_{2}^2 Q_2$, and two-reservoir Hamiltonian $H_B=H_{B1}+H_{B2}$.  At ultrastrong coupling  we find the \MFstate~of the combined system $S$ to be, see \App~\ref{sub:twoqubits}, 
\begin{equation} \label{eq:twobaths}
	\rho_S=\frac{e^{-\beta\sum_{mn}P_{1m} \otimes P_{2n} \, H_S \, P_{1m}\otimes P_{2n}}}{\tr\left[ e^{-\beta\sum_{m'n'}P_{1m'}  \otimes P_{2n'} \, H_S \, P_{1m'} \otimes P_{2n'}} \right]},
\end{equation}
where  $P_{\alpha n}=|x_{\alpha n}\rangle\langle x_{\alpha n}|$ are projection operators on the eigenstates  $|x_{\alpha n}\rangle$ of $X_{\alpha}$ for $\alpha=1,2$. 
We note that for two coupled systems interacting with a \emph{common reservoir}~\cite{Orth2010,Benatti2010,Deng2016}, the same expression \eqref{eq:twobaths} will follow.
Beyond the one-dimensional baths considered here,  determining the \MFstate~for three-dimensional systems, such as a single spin coupled simultaneously to baths in three dimensions \cite{Anders2020}, require multi-bath extensions of \eqref{eq:NormalizedMFGSmain}. It would also be interesting to establish  what state the system would take, if it is in contact with two reservoirs at different temperatures.

\medskip

As an example for Eq.~\eqref{eq:twobaths} we consider a {\bf Two qubit} system with Hamiltonian $H_1+H_2+H^{int}_{12}=\tfrac{1}{2}\omega_q\left(\sigma_{1z}+\sigma_{2z}\right)+\lambda_S\left(\sigma_{1+}\sigma_{2-}+\sigma_{1-}\sigma_{2+}\right)$ with $\sigma_{nz}=|e_n\rangle\langle e_n|-|g_n\rangle\langle g_n|$ and $\sigma_{n-}=|g_n\rangle\langle e_n|=\sigma^{\dag}_{n+}$ for $n=1,2$ and $\lambda_S$ the inter-qubit coupling. The qubits are each ultrastrongly coupled to a reservoir in $x$-direction, i.e. $X_1 =\sigma_{1x}$ and $X_2 = \sigma_{2x}$ with $\sigma_{nx}=\sigma_{n+}+\sigma_{n-}$. The required projection operators are then $P_{n\pm}=|\pm_n\rangle\langle \pm_n|$ with $|\pm_n\rangle=\left(|e_n\rangle\pm|g_n\rangle\right)/\sqrt{2}$. Thus, at ultrastrong coupling, the \MFstate~of the two qubits is, see \App~\ref{sub:twoqubits},
\begin{equation}\label{eq:twoqubitMFGS_JC}
	\rho_S=\frac{1}{4}\left(1-\sigma_{1x} \otimes \sigma_{2x} \tanh(\tfrac{1}{2}\beta\lambda_S)\right),
\end{equation}
which is independent of $\omega_q$. 
Clearly, while the system-environment coupling parameter $\lambda$ does not feature, the state does depend on the inter-qubit coupling $\lambda_S$. For vanishing $\lambda_S$ the state loses its energetic coherences, while at high $\lambda_S$ these are maximized.

For comparison, the conjectured steady state  \cite{Goyal2019} 
$\varrho_S^{conj}=\frac{1}{4}\left(1-\sigma_{1x} \otimes \sigma_{2x}\frac{\sinh\beta\lambda_S}{\cosh\beta\omega_q+\cosh\beta\lambda_S}\right)$  also has tensor product structure, but depends on $\omega_q$. It looses this dependence and becomes identical to \eqref{eq:twoqubitMFGS_JC}  at high temperatures, $\beta \omega_q \ll 1$. Dynamical convergence to $\varrho_S^{conj}$ was numerically evidenced using HEOM  \cite{Goyal2019}, but is again consistent with convergence to \eqref{eq:twoqubitMFGS_JC} for the parameters chosen, see figure in \App~\ref{sub:twoqubits}. 

Future research could provide a clearer disambiguation at lower temperatures, however, the HEOM method has its own convergence restrictions that may limit the range of parameters that can be explored. {A recent alternative numerical method, TEMPO, is based on time-evolving matrix product operators \cite{Strathearn2018}, and can efficiently describe the time evolution of quantum systems coupled to a non-Markovian harmonic environment. Furthermore, analytical approaches to solve the dynamics when the coupling is no longer weak could be based on reaction coordinate methods \cite{Burkey1984,Garg1985,Iles-Smith2014,Iles-Smith2016,Strasberg2016,Newman2017,Schaller2018,Nazir2018,Strasberg2018b,Restrepo2018,Anto-Sztrikacs2021}.}

\medskip

\emph{Conclusion and open questions.~} 
For general quantum systems coupled to a bosonic reservoir two explicit expressions for the \MFstate~$\rho_S$ defined in Eq.~\eqref{eq:DefMFGS} have been derived. 
Results \eqref{eq:PartitionedState} and \eqref{eq:NormalizedMFGSmain}  are valid for any quantum system, be it single qubits, V-systems, harmonic oscillators or others, and make it possible to immediately identify $\rho_S$ for a multitude of problems and applications arising in quantum thermodynamics, quantum thermometry, magnetism, quantum biology, and beyond.
Taken together,  expressions \eqref{eq:NormalizedMFGSmain} and \eqref{eq:PartitionedState} evidence the departure from the text book \Gib~state, diagonal in $H_S$, whenever the system-environment coupling is not negligible and the temperature is not large. With increasing coupling, the basis $\{P_n\}$ of the system's interaction operator $X$ becomes increasingly dominant, culminating  in the ultrastrong limit where it is the \emph{only} relevant basis.

To further explore this transition will require finding \MFstate~expressions for intermediate coupling strengths --- a very difficult analytical problem that may first be solved with numerical methods. We highlight that, while it is known that the dynamics of an open quantum system converges to the \MFstate~in the weak coupling limit  \cite{Mori2008, Fleming2011,Thingna2012,Subasi2012,Purkayastha2020}, the jury is out for the intermediate and ultrastrong coupling limit.
Another open question concerns extensions of the \MFstate~forms derived here to systems coupled to multiple bosonic reservoirs at the same temperature \cite{Anders2020}, besides the two qubit example given in Eq.~\eqref{eq:twoqubitMFGS_JC}.  Finally, we expect that extensions to fermionic reservoirs can readily be made, as the  derivations leading to  \eqref{eq:NormalizedMFGSmain} and \eqref{eq:PartitionedState} do not centrally depended on the bosonic nature of the reservoir.

\smallskip

{\bf Note added:} A paper has recently been posted \cite{Trushechkin2021} that derives an ultrastrong coupling master equation and confirms dynamical convergence to the ultrastrong \MFstate~\eqref{eq:PartitionedState} derived here.

\begin{acknowledgments}
We thank Ryoichi Kawai, Steve Barnett, Marco Berritta, Daniel Burgarth, Federico Cerisola, Luis Correa, Giacomo Guarnieri, Simon Horsley and Stefano Scali for inspiring discussions.  JA and JDC acknowledge funding from EPSRC (EP/R045577/1), and JA thanks the Royal Society for support.
\end{acknowledgments}

\bibliographystyle{apsrev4-1.bst}
\bibliography{HMFBib.bib}



\onecolumngrid

\newpage

\noindent {``Weak and ultrastrong coupling limits of the quantum mean force \Gib~state''},  J. D. Cresser, J. Anders

\bigskip
\bigskip

\appendix
\begin{center} 

 {\bf \Large  Appendix} 
 
\end{center}

\section{High temperature limit via a factorization approximation\label{app:factorization}}

The high temperature limit corresponds to $\beta\to0$.  We make use of the product approximation obtained via the Zassenhaus formula \cite{Suzuki1977} to write it in symmetrized form, with $H_0=H_S'+H_R$,
\begin{equation}
	e^{-\beta H_{SR}}
	= e^{-\beta(H_0+ \lambda V)}
	=\frac{1}{2}\left[e^{-\beta H_0}e^{-\beta  \lambda V}+e^{-\beta  \lambda V}e^{-\beta H_0} \right]\left(1+\mathcal{O}(\beta^2)\right).
\end{equation}
With interaction  $V = X \, B$ with  $B=\int_{0}^{\infty} \d \omega \, \sqrt{J(\omega)}\left(b(\omega)+b^\dagger(\omega)\right)$, and reservoir thermal state $\tau_R(\beta) ={e^{-\beta H_R} / \tr_R[e^{-\beta H_R}]} $, and denoting the un-normalized state with a tilde, one obtains the reduced state
\begin{equation}
	\tilde{\rho}_S
	=\left(e^{-\beta H_S'} \, \tr_R\left[ \tau_R(\beta) \, e^{-\beta \lambda XB}\right]
	+  \tr_R\left[ \tau_R(\beta) \, e^{-\beta \lambda XB}\right] \, e^{-\beta H_S'}\right)\left(1+\mathcal{O}(\beta^2)\right),
\end{equation}
where any {common} factors have been ignored as we ultimately will have to normalize this {state}. Further, given that $\tau_R(\beta)$ is a Gaussian state, we can immediately evaluate the trace over reservoir states to give
\begin{equation}\label{eq:GaussianAverage}
	\tr_R\left[ {\tau_R(\beta)} \, e^{-\beta {\lambda} XB}\right]
	=e^{\frac{1}{2}\beta^2 {\lambda^2} X^2\langle B^2\rangle}		
\end{equation}
where 
\begin{equation}
	\langle B^2\rangle
	=\int_{0}^{ \infty}\d\omega \, J(\omega)\coth \frac{\beta\omega}{2}.		
\end{equation}
Using $\coth(x)=x^{-1}+\mathcal{O}(x^2)$ for $x=\tfrac{1}{2}\beta\omega$, we have for small $\beta$, on noting that only the first term will contribute as the exponent is already second order in $\beta$, 
\begin{equation}
	\langle B^2\rangle
	=\frac{2}{\beta} \int_{0}^{\infty}\d\omega \, \frac{J(\omega)}{\omega}=\frac{2Q}{\beta},
\end{equation}
where $Q$ is the reorganization energy. Thus we have
\begin{equation}
	\tilde{\rho}_S=
	\left(e^{-\beta H_S'}e^{\beta {\lambda^2} X^2Q}+e^{\beta {\lambda^2} X^2Q}e^{-\beta H_S'}\right)\left(1+\mathcal{O}(\beta^2)\right),
\end{equation}
at which point we can reverse the product approximation for small $\beta$. Noting that $H_S'=H_S+{\lambda^2} X^2Q$ {and normalizing, one finds}  $\rho_S = \tau_S {+ \mathcal{O}(\beta^2)}$ i.e., the high temperature mean force state is just the \Gib~state, $\tau_S$ as stated in the main text. This result is valid for all coupling strengths such that the energy scale set by the temperature is larger than that of the system-reservoir coupling.

\section{Kubo expansion for weak coupling at arbitrary temperature}\label{app:ArbTempWeakCoup}

This is the approach used in Mori and Miyashita \cite{Mori2008}, Suba\c{s}{\i} \emph{et al.}, \cite{Subasi2012}, Thingna \emph{et al.}, \cite{Thingna2012}, and appeared earlier in Geva \emph{et al.} \cite{Geva2000}, and in the recent paper by Purkayastha \emph{et al.}, \cite{Purkayastha2020} and involves using the so-called Kubo expansion \cite{Kubo1957}, analogous to the corresponding expansion for the time evolution operator, albeit for imaginary time
\begin{equation}
	e^{-\beta(H_0+\lambda V)}=e^{-\beta H_0}
	\left(1-\lambda \int_{0}^{\beta}\d\beta_1e^{\beta_1H_0}Ve^{-\beta_1H_0}
	+\lambda^2\int_{0}^{\beta}\d\beta_2\int_{0}^{\beta_2}\d\beta_1e^{{+}\beta_2H_0}
	Ve^{-(\beta_2-\beta_1)H_0}Ve^{-\beta_1H_0}+\mathcal{O}(\lambda^3) \right).
\end{equation}

\subsection{Hermiticity of perturbative result}

In the case of the unitary time evolution operator, the expansion cut off at a finite number of terms is no longer unitary. However, here, the operator $\exp[-\beta H]$ {for $H=H_0 + \lambda V$} is Hermitian, and the finite order expansion is also {Hermitian} i.e., there is no requirement to symmetrize the expression, cf., \cite{Purkayastha2020}. To show this, we can write $e^{-\beta H}=e^{-\beta H_0}F(\beta)$ where
\begin{equation}
	F(\beta)=1-\lambda \int_{0}^{\beta}\d\beta_1e^{\beta_1H_0}Ve^{-\beta_1H_0}
	+\lambda^2\int_{0}^{\beta}\d\beta_2\int_{0}^{\beta_2}\d\beta_1e^{{+}\beta_2H_0}
	Ve^{-(\beta_2-\beta_1)H_0}Ve^{-\beta_1H_0}+\mathcal{O}(\lambda^3).
\end{equation}
This quantity is clearly {non-Hermitian}, but the product $e^{-\beta H_0}F(\beta)=e^{-\beta H}$ clearly is. The question is whether or not cutting off the expansion of $F(\beta)$ to second order breaks this Hermiticity. So we will write $F(\beta)=F^{(2)}(\beta)+\mathcal{O}(\lambda^3)$. We want to check whether or not the quantity $e^{-\beta H_0}F^{(2)}(\beta)$ is Hermitian. To do this, take the hermitian conjugate
\begin{equation}
	\left(e^{-\beta H_0}F^{(2)}(\beta)\right)^\dagger
	=e^{-\beta H_0}
	- {\lambda} \int_{0}^{\beta}\d\beta_1e^{-\beta_1H_0}Ve^{-(\beta-\beta_1)H_0}
	+{\lambda^2} \int_{0}^{\beta}\d\beta_2\int_{0}^{\beta_2}\d\beta_1
	e^{-\beta_1H_0}Ve^{-(\beta_2-\beta_1)H_0}Ve^{-(\beta-\beta_2)H_0}.
\end{equation}
Now make the changes of variables $\beta_1'=\beta-\beta_1$ in the first integral, and $\beta_1'=\beta-\beta_2$ and $\beta-\beta_1=\beta_2'$ in the second integral. We immediately get $\left(e^{-\beta H_0}F^{(2)}(\beta)\right)^\dagger=e^{-\beta H_0}F^{(2)}(\beta)$.

\subsection{Perturbative result for general system operator $X$}

We now want to expand the Mori and Miyashita result making it more explicit for practical application. Using $V=XB$, $H_0=H_S'+H_R$ where $H_S'=H_S+\lambda^2X^2Q$, we can define, in a `$\beta$-interaction picture'	$V(-i\beta)=e^{\beta H_0}Ve^{-\beta H_0}=X(-i\beta)B(-i\beta)$. We then find, recognizing that ${\tau}_R(\beta)=e^{-\beta H_R}/Z_R$ and on taking the trace over $R$, that the unnormalized state of the system (indicated with a tilde, $\tilde{\rho}_S$)
\begin{equation}\label{eq:MFGSunnormal}
	\tilde{\rho}_S = {\tr_R}\left[e^{-\beta H_{SR}}\right]
	= {e^{-\beta H_S'} }
	\left(1+\lambda^2\int_{0}^{\beta}\d\beta_2\int_{0}^{\beta_2}\d\beta_1X(-i\beta_2)X(-i\beta_1)G(\beta,-i(\beta_2-\beta_1))+\mathcal{O}(\lambda^4)\right),
\end{equation}
where we have used $\tr_R\left[B \, {\tau}_R(\beta)\right]=0$ and where 
\begin{equation}\label{eq:correlationFunctionWrtBeta}
	G(\beta,-i(\beta_2-\beta_1))=\tr_R\left[B(-i\beta_2)B(-i\beta_1) \, {\tau}_R(\beta)\right],
\end{equation} 
is the $\beta$-analogue of the temporal reservoir correlation function. 

This result \eqref{eq:MFGSunnormal} is useful for finding the high temperature limit $\beta\to0$, and is discussed later in Section \ref{sec:HighTempLimitWeakCoupling}, but as it stands, this result is not in a form for practical application. To achieve this necessitates rewriting $X$ in terms of the energy eigenoperators for an arbitrary multilevel system. 

\medskip

The system operator $X$ can be expressed in terms of the energy eigenoperators for the system 
\begin{equation}
	X=\sum_{n=-N}^{+N}X_n
\end{equation}
where $\left[H_S,X_n\right]=\omega_nX_n$, and where the $\omega_n$ are Bohr frequencies, where $m$ ranges over $2N+1$ positive and negative values, and $\omega_n=-\omega_{-n}$. Finally, since $X=X^\dagger$ we have $X_n=X_{-n}^\dagger$, and $\omega_0=0$, though in general, $X_0\ne 0$. 

The role of the $X_n$ can be seen by considering, for an eigenstate $|\omega\rangle$ of $H_S$, $\left[H_S,X_n\right]|\omega\rangle=H_S\left(X_n|\omega\rangle\right)-\omega X_n|\omega\rangle=\omega_nX_n|\omega\rangle$ so $H_S(X_n|\omega\rangle)=(\omega+\omega_n)X_n|\omega\rangle$. Provided $X_n|\omega\rangle\ne 0$, the effect of $X_n$ on the state $|\omega\rangle$ is to map it to a new state $X_n|\omega\rangle$ with energy $\omega+\omega_n$. So, if $n>0$, $X_n$ is a raising operator, while if $n<0$ it is a lowering operator.
We then have
\begin{equation}\label{eq:expXexp}
	e^{\beta H_S'}Xe^{-\beta H_S'}=X(-i\beta)=e^{\beta H_S}Xe^{-\beta H_S}+\mathcal{O}(\lambda^2)=\sum_{n}X_n e^{\beta\omega_n}+\mathcal{O}(\lambda^2).
\end{equation}
Note that here and in the following, the sum will be over the $2N+1$ integers $-N$ to $N$, but the limits of the sum will be implied.

Feeding \eqref{eq:expXexp} into expression (\ref{eq:MFGSunnormal}) we get, correct to second order, for the unnormalized density operator
\begin{equation}
	\tilde{\rho}^{\,(2)}_S=e^{-\beta H_S'}+\lambda^2e^{-\beta H_S}
	\sum_{mn}X_mX_n^\dagger\int_{0}^{\beta}\d\beta_2\int_{0}^{\beta_2}\d\beta_1 \,
	e^{\beta_2\omega_m}e^{-\beta_1\omega_n}G(\beta,-i(\beta_2-\beta_1)).
\end{equation}
A simple change of variable, and swapping the order of integration puts this in the form
\begin{equation}
	\tilde{\rho}^{\,(2)}_S=e^{-\beta H_S'}+\lambda^2e^{-\beta H_S}
	\sum_{mn}X_mX_n^\dagger\int_{0}^{\beta}\d\beta_1\int_{\beta_1}^{\beta}\d\beta_2 \,
	e^{\beta_2\omega_{mn}}e^{\beta_1\omega_n}G(\beta,-i\beta_1),
\end{equation}
{with $\omega_{mn}=\omega_m - \omega_n$}, which after carrying out the $\beta_2$ integral gives
\begin{equation}
	\tilde{\rho}^{\,(2)}_S=e^{-\beta H_S'}+\lambda^2e^{-\beta H_S}
	\sum_{mn}\omega_{mn}^{-1}X_mX_n^\dagger\int_{0}^{\beta}\d\beta_1\left(e^{\beta\omega_{mn}}e^{\beta_1\omega_n}-e^{\beta_1\omega_{m}}\right)
	G(\beta,-i\beta_1).
\end{equation}

Appearing here is the correlation function $G(\beta,-i\beta)$ defined in \eqref{eq:correlationFunctionWrtBeta}, which is the $t\to-i\beta$ form of the temporal correlation function $G(\beta,t)=\tr[\tau_R(\beta)B(t)B(0)]$ where
\begin{equation}
	B(t)=\int_0^\infty \d\omega\sqrt{J(\omega)}\left(b(\omega)e^{-i\omega t}+b^\dagger(\omega)e^{i\omega t}\right)
\end{equation}
from which it readily follows that
\begin{equation}
	G(\beta,t)=\int_0^\infty \d\omega J(\omega)\left((n_\beta(\omega)+1)e^{-i\omega t}+n_\beta(\omega)e^{i\omega t}\right)
\end{equation}
with $n_{\beta}(\omega)=1/(e^{\beta\omega}-1)$. Hence for the correlation function $G(\beta,-i\beta_1)$ we have
\begin{equation}\label{eq:CorrFuncBeta}
	G(\beta,-i\beta_1)=\int_{0}^{\infty}\d\omega J(\omega)\left((n_\beta(\omega)+1)e^{-\beta_1\omega}+n_\beta(\omega)e^{\beta_1\omega}\right).
\end{equation}
An integral transform of this,
\begin{equation}
	\int_{0}^{\beta}\d\beta_1 \, e^{\beta_1\omega_n}G(\beta,-i\beta_1)
	=A_\beta(\omega_n)+e^{\beta\omega_n}A_\beta(-\omega_n),
\end{equation}
defines the function $A_\beta(\omega_n)$, generally to be understood as a principal part integral, given by
\begin{equation}\label{eq:AbetaOmega}
	A_\beta(\omega_n)=\int_{0}^{\infty}\d\omega J(\omega)\left(\frac{n_\beta(\omega)+1}{\omega-\omega_n}-\frac{n_\beta(\omega)}{\omega+\omega_n}\right)
	=\int_{0}^{\infty}\d\omega \, J(\omega)\left(\frac{\omega_n\coth(\tfrac{1}{2}\beta\omega)}{\omega^2-\omega_n^2}+\frac{\omega}{\omega^2-\omega_n^2}\right),
\end{equation}
where we have used $2n_\beta(\omega)+1=\coth(\tfrac{1}{2}\beta\omega)$. This expression can be recognized as appearing in the second order energy level `Lamb shift' contribution to the system Hamiltonian, given by $\Delta H_{LS}=-\sum_{n}A_\beta(\omega_n)X_nX_n^\dagger$.
We can now write
\begin{equation}
	\tilde{\rho}^{\,(2)}_S=e^{-\beta H_S'}+\lambda^2e^{-\beta H_S}
	\sum_{mn}\omega_{mn}^{-1}X_mX_n^\dagger\left(e^{\beta\omega_{mn}}A_\beta(\omega_n)-A_\beta(\omega_m)+e^{\beta\omega_m}\left(A_\beta(-\omega_n)-A_\beta(-\omega_m)\right)\right).
\end{equation}
It is convenient at this stage to separate out the $m=n$ contribution to the double sum, so we have
\begin{eqnarray}\label{eq:XmXnDagger}
		\tilde{\rho}_S^{(2)}
		=e^{-\beta H_S'}
		&+&\lambda^2e^{-\beta H_S}\sum_{n}X_nX_n^\dagger
		\left(\beta A_\beta(\omega_n)-\frac{dA_\beta(\omega_n)}{d\omega_n}
		+e^{\beta\omega_n}\frac{dA_\beta(\omega_{-n})}{d\omega_{-n}}\right) \\
		&+&\lambda^2e^{-\beta H_S}\sum_{m\ne n}\omega_{mn}^{-1}X_mX_n^\dagger
		\left[e^{\beta\omega_{mn}}A_\beta(\omega_n)+e^{\beta\omega_m}A_\beta(-\omega_n)-\left(A_\beta(\omega_m)+e^{\beta\omega_m}A_\beta(-\omega_m)\right)\right], \nonumber 
\end{eqnarray}
where $\sum_{m \neq n}$ is a sum over all ordered pairs $(m, n)$ with $m\neq n$.
The exponential factors appearing in Eq.\ (\ref{eq:XmXnDagger}) can now be removed by use of $e^{-\beta H_S}X_m=X_me^{-\beta H_S}e^{-\beta \omega_m}$ so that
\begin{equation}\label{eq:swappingOperators}
	e^{-\beta H_S}X_mX_n^\dagger=X_mX_n^\dagger e^{-\beta H_S}e^{-\beta\omega_{mn}}.
\end{equation}
This yields
\begin{eqnarray}
		\tilde{\rho}_S^{\,(2)}
		=e^{-\beta H_S'}
		&+&\lambda^2\sum_{n}
		\left(e^{-\beta H_S}\beta X_nX_n^\dagger A_\beta(\omega_n)
		-e^{-\beta H_S}X_nX_n^\dagger\frac{dA_\beta(\omega_n)}{d\omega_n}
		+X_ne^{-\beta H_S}X_n^\dagger \frac{dA_\beta(\omega_{-n})}{d\omega_{-n}}\right) \\
		&+&\lambda^2\sum_{m\ne n}\omega_{mn}^{-1}
		\left(X_mX_n^\dagger e^{-\beta H_S}A_\beta(\omega_n)
		-e^{-\beta H_S}X_mX_n^\dagger A_\beta(\omega_m)+X_me^{-\beta H_S}X_n^\dagger
		\left(A_\beta(-\omega_n)-A_\beta(-\omega_m)\right)\right). \nonumber
\end{eqnarray}
We now relabel the summation $n\to-n,m\to-m$ indices, so that $A_\beta(\omega_{-n})\to A_\beta(\omega_n)$, and similarly for the derivatives, recalling when doing so that the sums are symmetric from $-N$ to $N$, that $\omega_{mn}\to-\omega_{mn}$ and $X_{-n}=X_n^\dagger$, to give
\begin{eqnarray}
		\tilde{\rho}_S^{\,(2)}
		=e^{-\beta H_S'}
		&+&\lambda^2\sum_{n}
		\left(e^{-\beta H_S}\beta X_nX_n^\dagger A_\beta(\omega_n)
		-\left(e^{-\beta H_S}X_nX_n^\dagger
		-X_n^\dagger e^{-\beta H_S}X_n\right)
		\frac{dA_\beta(\omega_n)}{d\omega_n}\right) \\
		&+&\lambda^2\sum_{m\ne n}\omega_{mn}^{-1}
		\left(X_mX_n^\dagger e^{-\beta H_S}A_\beta(\omega_n)
		-e^{-\beta H_S}X_mX_n^\dagger A_\beta(\omega_m)-X_m^\dagger e^{-\beta H_S}X_n
		\left(A_\beta(\omega_n)-A_\beta(\omega_m)\right)\right). \nonumber
\end{eqnarray}
Next, relabel the indices $m\leftrightarrow n$ for all terms involving $A_\beta(\omega_m)$, with once again $\omega_{mn}\to-\omega_{mn}$, giving
\begin{eqnarray}
		\tilde{\rho}_S^{\,(2)}
		=e^{-\beta H_S'}
		&+&\lambda^2\sum_{n}
		\left(e^{-\beta H_S}\beta X_nX_n^\dagger A_\beta(\omega_n)
		+\left[X_n^\dagger, e^{-\beta H_S}X_n\right]
		\frac{dA_\beta(\omega_n)}{d\omega_n}\right)\\
		&+&\lambda^2\sum_{m\ne n}\omega_{mn}^{-1}
		\left(X_mX_n^\dagger e^{-\beta H_S}
		+e^{-\beta H_S}X_nX_m^\dagger -X_m^\dagger e^{-\beta H_S}X_n
		-X_n^\dagger e^{-\beta H_S}X_m\right)A_\beta(\omega_n) . \nonumber
\end{eqnarray}
Regrouping everything into commutators and using $\tau_S=e^{-\beta H_S}/Z_S$ then gives
\begin{eqnarray}\label{eq:generalResult}
		\tilde{\rho}_S^{(2)}
		=e^{-\beta H_S'}
		&+&\lambda^2Z_S\sum_{n}
		\left(\beta \tau_SX_nX_n^\dagger A_\beta(\omega_n)
		+\left[X_n^\dagger, \tau_SX_n\right]
		\frac{dA_\beta(\omega_n)}{d\omega_n}\right) \\
		&+&\lambda^2Z_S\sum_{m\ne n}\omega_{mn}^{-1}
		\left(\left[X_m,X_n^\dagger \tau_S\right]
		+\left[\tau_SX_n,X_m^\dagger\right]\right)A_\beta(\omega_n) . \nonumber
\end{eqnarray}

It could be argued that $e^{-\beta H_S'}$ with $H_S'=H_S+\lambda^2X^2Q$ could be evaluated exactly, at least for low dimensional systems, but given that that is not necessarily the case, we will proceed to obtaining an expansion to second order in the interaction, due to the $X^2 Q$ contribution, of $e^{-\beta H_S'}$.  This requires use of the Kubo expansion again, to give
\begin{equation}\label{eq:tauSdashExpansion}
	e^{-\beta H_S'}=e^{-\beta H_S}-\lambda^2Q\sum_{m\ne n}\omega_{mn}^{-1}\left[X_mX_n^\dagger, e^{-\beta H_S}\right]-\lambda^2Q\beta\sum_{n}e^{-\beta H_S}X_nX_n^\dagger +\mathcal{O}(\lambda^4).
\end{equation}
By a suitable renaming of summation indices we have 
\begin{equation}	
\sum_{m\ne n}\omega_{mn}^{-1}\left[X_mX_n^\dagger, e^{-\beta H_S}\right]
	=\sum_{m\ne n}\omega_{mn}^{-1}\left(\left[X_m,X_n^\dagger e^{-\beta H_S}\right]
	+\left[e^{-\beta H_S}X_n,X_m^\dagger\right]\right)
\end{equation}
which, upon substitution into \eqref{eq:generalResult}, combines with a similar term in appearing there to give, on dropping the common factor of $Z_S$
\begin{eqnarray}
		\tilde{\rho}_S^{(2)}
		=\tau_S
		&+&\lambda^2\sum_{n}
		\left(\beta \tau_SX_nX_n^\dagger \D_\beta(\omega_n)
		+\left[X_n^\dagger, \tau_SX_n\right]
		\frac{dA_\beta(\omega_n)}{d\omega_n}\right) \\
		&+&\lambda^2\sum_{m\ne n}\omega_{mn}^{-1}
		\left(\left[X_m,X_n^\dagger \tau_S\right]
		+\left[\tau_SX_n,X_m^\dagger\right]\right)\D_\beta(\omega_n), \nonumber
\end{eqnarray}
where  we have introduced   
\begin{equation}
	\D_\beta(\omega_n)
	=
	\begin{cases}
	A_\beta(\omega_n)-Q & \text{if }X^2 {\not \propto} \,  \mathbb{I}\\
	A_\beta(\omega_n) & \text{if }X^2 {\propto} \, \mathbb{I}
	\end{cases}
\end{equation}
with the simplification in second line occurring since $\tau_S'\to\tau_S$ for $X^2 { \propto}  \,\mathbb{I}$, as in this case, the $X^2Q$ contribution represents a constant energy off-set.

\subsection{Normalization}\label{sec:Normalization}

Normalization requires calculating the trace of $\tilde{\rho}_S^{\,(2)}$. We find
\begin{equation}
	Z_S^{(2)}=\tr\left[\tilde{\rho}_S^{\,(2)}\right]=
	1+\beta\lambda^2\sum_{n} \tr_S\left[\tau_SX_nX_n^\dagger\right]\D_\beta(\omega_n).
\end{equation}
Then provided 
\begin{equation}\label{eq:binomialApproxCondition}
	{\left| \beta\lambda^2\sum_{n}\tr_S\left[\tau_SX_nX_n^\dagger\right]\D_\beta(\omega_n) \right|}  \ll1,
\end{equation}
which defines an upper limit to $\beta$, we can use the binomial approximation and expand
\begin{equation}\label{eq:binomialApprox}
	\frac{1 }{ Z_S^{(2)}}=1-\beta\lambda^2\sum_{n}\tr_S\left[\tau_SX_nX_n^\dagger\right]\D_\beta(\omega_n).
\end{equation}
This leads to the following expression, correct to second order in the interaction, for the normalized density operator
\begin{eqnarray}\label{eq:NormalizedMFGS}
		\rho_S^{(2)}=\frac{\tilde{\rho}_S^{(2)} }{ Z_S^{(2)}}
		=\tau_S
		&+&\lambda^2\sum_{n}
		\left[X_n^\dagger, \tau_SX_n\right]
		\frac{d{\D}_\beta(\omega_n)}{d\omega_n}
		+\lambda^2\beta\sum_{n}\tau_S\left(X_nX_n^\dagger
		- \tr_S\left[\tau_SX_nX_n^\dagger\right]\right)\D_\beta(\omega_n) \nonumber \\
		&+&\lambda^2\sum_{m\ne n}\omega_{mn}^{-1}
		\left(\left[X_m,X_n^\dagger \tau_S\right]
		+\left[\tau_SX_n,X_m^\dagger\right]\right)\D_\beta(\omega_n). 
\end{eqnarray}
This is {result Eq.~\eqref{eq:NormalizedMFGSmain}} in the main text.

\subsection{High temperature limit}\label{sec:HighTempLimitWeakCoupling}

In the high temperature limit $\beta\to0$, it is easiest to work with the earlier form \eqref{eq:MFGSunnormal}, this requiring an expansion to second order in $\beta$, from which we find
\begin{equation}
	\tilde{\rho}_S^{\,(2)}= e^{-\beta H_S'}\left(1+\beta\lambda^2 X^2Q\right).
\end{equation}
Expanding $e^{-\beta H_S'}$ to second order in the interaction in a similar manner to \eqref{eq:MFGSunnormal} (or using \eqref{eq:tauSdashExpansion}) and taking the high temperature limit $\beta\to0$ yields
\begin{equation}
	e^{-\beta H_S'}=e^{-\beta(H_S+\lambda^2X^2Q)}=e^{-\beta H_S}
	-\lambda^2\int_{0}^{\beta}\d\beta'e^{-(\beta-\beta')H_S}X^2Qe^{-\beta' H_S}
	\sim e^{-\beta H_S}\left(1-\beta \lambda^2X^2Q\right),
\end{equation}
so that to second order we get
\begin{equation}
	\tilde{\rho}_S^{\,(2)}= e^{-\beta H_S}\left(1-\beta \lambda^2X^2Q\right)\left(1+\beta\lambda^2 X^2Q\right)=e^{-\beta H_S}+\mathcal{O}(\lambda^4).
\end{equation}
Hence, here the reorganization energy $\beta\lambda^2X^2Q$ is cancelled and we find $\tilde{\rho}_S^{\,(2)}= e^{-\beta H_S}$ to second order, which is just the result obtained by the factorization method.

Alternatively, and as a sanity check, the same limit can be taken for \eqref{eq:NormalizedMFGS}. This involves expanding all the $\lambda$-dependent terms in \eqref{eq:NormalizedMFGS} in powers of $\beta$. The term with the prefactor $\beta^{-1}$ is found to vanish in the limit of $\beta\to0$. Constant terms (i.e., those independent of $\beta$) also cancel, as do all those terms proportional to $Q$. The remaining terms, all of higher order in $\beta$ will then vanish in the limit of $\beta \to 0$. What is left in the high temperature limit is thus once again $e^{-\beta H_S}$.

\subsection{Low temperature limit}\label{App:WeakCouplingLowTempLimit}

In the low temperature limit, $\beta\omega_S\gg1$, (while assuming \eqref{eq:binomialApproxCondition} is fulfilled) where $\omega_S$ is a typical Bohr frequency for the system, in Eq.~\eqref{eq:NormalizedMFGS} we can replace $\tau_S\to|0\rangle\langle 0|$ where $|0\rangle$ is the ground state of the system, assumed to be non-degenerate. This gives
\begin{eqnarray}\label{eq:lowTAppendix}
		\lowT =|0\rangle\langle 0|
		&+&\lambda^2\sum_{n}
		\left[X_n^\dagger, |0\rangle\langle 0|X_n\right]
		\frac{d  {\D}_\beta(\omega_n)}{d\omega_n}
		+\lambda^2\beta\sum_{n}|0\rangle\langle 0|\left(X_nX_n^\dagger-
		\, \tr_S\left[|0\rangle\langle 0|X_nX_n^\dagger\right]\right)  {\D}_\beta(\omega_n) \nonumber \\
		&+&\lambda^2\sum_{m\ne n}\omega_{mn}^{-1}
		\left(\left[X_m,X_n^\dagger |0\rangle\langle 0|\right]
		+\left[|0\rangle\langle 0|X_n,X_m^\dagger\right]\right) {\D}_\beta(\omega_n).\end{eqnarray}
The problem terms are the potentially divergent terms proportional to $\beta$. Separating out the term which includes this contribution, we have to deal with the term $\langle 0|X_nX_n^\dagger|0\rangle\langle 0|-\langle 0|X_nX_n^\dagger  \equiv \langle \psi|$ where we have defined an unnormalised state $|\psi \rangle$. 
Since $\left[H_S,X_nX_n^\dagger\right]=0$, and denoting the ground state energy as $E_0$, i.e. $H_S|0\rangle=E_0 |0\rangle$ with $|0\rangle$ the ground state of the system, one has 
\begin{equation}
	\langle \psi| H_S 
	= \langle 0|X_nX_n^\dagger|0\rangle\langle 0|H_S -\langle 0|X_nX_n^\dagger  H_S 
	= E_0 \langle 0|X_nX_n^\dagger|0\rangle\langle 0|  - E_0 \langle 0| X_nX_n^\dagger   
	= E_0 \langle \psi|.
\end{equation}
Since the ground state is non-degenerate we thus must have $ \langle \psi| = p  \langle 0|$ for some c-number $p$. But from the definition we have  $ \langle \psi| 0 \rangle = 0$ and hence $p=0$. So $\langle \psi| =0$, i.e. $\langle 0|X_nX_n^\dagger|0\rangle\langle 0|-\langle 0|X_nX_n^\dagger =0$. The terms linear in $\beta$ in \eqref{eq:lowTAppendix} cancel and the result converges. 

Further, we can note that a structure like $X_n|0\rangle$ will only be non-zero for $n\ge0$, so the frequencies $\omega_n$ appearing here will always be negative (or zero). It is then useful to make a notational change $n\to-n, m\to-m$ so that $\omega_n\to\omega_{-n}=-\omega_n$. We can also put $n_\beta(\omega)=0$ in the expression for $A_\beta(\omega_n)$ and so are then left with
\begin{eqnarray}\label{eq:LowTempMFGS}
		\lowT 
		=|0\rangle\langle 0|
		&+& \lambda^2\sum_{n}
		\left[X_n, |0\rangle\langle 0|X_n^\dagger\right]
		\int_{0}^{\infty} \d\omega \, \frac{J(\omega)}{(\omega+\omega_n)^2} \nonumber \\
		&+&\lambda^2\sum_{m\ne n}\omega_{mn}^{-1}
		\left(\left[X_n |0\rangle\langle 0|,X_m^\dagger\right]
		+\left[X_m,|0\rangle\langle 0|X_n^\dagger\right]\right)\left(\int_{0}^{\infty} \d\omega \, \frac{J(\omega)}{\omega+\omega_n}-Q\right) .
\end{eqnarray}
as the leading terms in the low temperature expression for $\rho_S^{(2)}$. The contribution explicitly involving $Q$ here is due to the reorganization energy term in $H_S'$, and will only be present if $X^2$ is not a multiple of the identity. If it is a multiple of the identity, then this $Q$ term can be dropped.

We remark that the $T=0$ result is not necessarily implied by $\lowT$ because of the need to satisfy the lower limit condition \eqref{eq:binomialApproxCondition}. Nevertheless, an alternate perturbation theory approach for $T=0$ shows that \eqref{eq:LowTempMFGS} is indeed the zero temperature result. This can be done by standard perturbation methods based on the idea (see also \cite{Subasi2012}) that in the absence of any interaction between the system and reservoir, the ground state (i.e., the zero temperature state) of the combined system would be just the state $|\phi^{(0)}\rangle=|0\rangle\otimes|\text{vac}\rangle$. Doing so leads to \eqref{eq:LowTempMFGS}.

\section{{Applications of the weak coupling result \eqref{eq:NormalizedMFGS}}}

\subsection{Weak coupling \MFstate~for a single qubit} \label{App:spinboson}

To provide an example for the general weak coupling \MFstate~given in Eq.\ (\ref{eq:NormalizedMFGS}), we consider the {\bf single qubit} coupled to a boson bath which has previously been discussed in \cite{Purkayastha2020}. Here the system Hamiltonian is  $H_S=\tfrac{1}{2}\omega_q\sigma_z$ and the coupling operator is $X=\cos\theta\sigma_z-\sin\theta\sigma_x$. This coupling enables energy exchange between the qubit and reservoir via $\sigma_x$ as well as dephasing in the energy basis $\sigma_z$. The energy eigenoperators can be readily identified, and we have
\begin{eqnarray} 
		X_1&=&-\sin\theta\sigma_+,\qquad \omega_1=+\omega_q \nonumber \\
		X_0 &=& + \cos\theta\sigma_z, \qquad\omega_0=0 \label{eq:xsingle} \\
		X_{-1}&=&-\sin\theta\sigma_-,\qquad \omega_{-1}=-\omega_q. \nonumber
\end{eqnarray}
Substitution of these expressions into the general result \eqref{eq:NormalizedMFGSmain}, with $\mathcal{D}_\beta(\omega_n)\to A_\beta(\omega_n)$ as $X^2=\mathbb{I}$ then gives, with $\tau_n=\langle n|\tau_S|n\rangle, n=0,1$
\begin{equation}\label{eq:Purkyrho}
	\begin{split}
		\rho_S^{(2)}&=\tau_S-\lambda^2 \,  \frac{\sin2\theta }{ \omega_q} \left[\tau_0A_\beta(-\omega_q)+\tau_1A_\beta(\omega_q)-A_\beta(0)\right]\sigma_x\\
		&\phantom{=}+\lambda^2\sin^2\theta\left[-\tau_0\frac{dA_\beta(-\omega_q)}{d\omega_q}-\tau_1\frac{dA_\beta(\omega_q)}{d\omega_q}+\beta\tau_1\tau_0\left(A_\beta(\omega_q)-A_\beta(-\omega_q)\right)\right]\sigma_z.
	\end{split}
\end{equation}

We can now make use of the expression \eqref{eq:AbetaOmega} for $A_\beta(\omega_n)$ in \eqref{eq:Purkyrho}. We further have $\tau_0=\tfrac{1}{2}\sech(\tfrac{1}{2}\beta\omega_q)e^{\frac{1}{2}\beta\omega_q}$ and $\tau_1=\tfrac{1}{2}\sech(\tfrac{1}{2}\beta\omega_q)e^{-\frac{1}{2}\beta\omega_q}$. Setting $\langle \sigma_z\rangle_0=\tau_1-\tau_0= - \tanh(\tfrac{1}{2}\beta\omega_q)$ as the bare system inversion, and substituting these last expressions into \eqref{eq:Purkyrho} then gives
\begin{eqnarray}
		\rho_S^{(2)}
		=\tau_S+\frac{\langle \sigma_x\rangle}{2}  \, \sigma_x
		+\frac{\langle \sigma_z\rangle-\langle \sigma_z\rangle_0}{2} \, \sigma_z,
\end{eqnarray} 
where the coefficient functions are 
\begin{eqnarray}\label{eq:sigmaX}
	\langle \sigma_x\rangle
	&=&2\lambda^2 \, \frac{\sin2\theta}{\omega_q}
		\left[ \tanh\left(\frac{\beta\omega_q }{ 2}\right) \, 
		\int_{0}^{\infty}\d\omega \, J(\omega) \, \coth\left(\frac{\beta\omega }{ 2}\right) \,
		\frac{\omega_q}{\omega^2-\omega_q^2}\right.\nonumber\\
		&&\hspace{2cm}\left.-\int_{0}^{\infty}\d\omega \, J(\omega)\frac{\omega}{\omega^2-\omega_q^2}+\int_{0}^{\infty}d\omega \frac{J(\omega)}{\omega}\right],
\end{eqnarray}
and
\begin{eqnarray}\label{eq:sigmaZ}
	\langle \sigma_z\rangle - \langle \sigma_z\rangle_0&=&
	2\lambda^2\sin^2\theta \, \left[ \tanh\left(\frac{\beta\omega_q }{ 2}\right) \,  
	\int_{0}^{\infty}\d\omega \, J(\omega) \, \coth\left(\frac{\beta\omega }{ 2}\right) \,
	\frac{\omega^2+\omega_q^2}{(\omega^2-\omega_q^2)^2}
	-\int_{0}^{\infty}\d\omega \, J(\omega) \, \frac{2\omega\omega_q}{(\omega^2-\omega_q^2)^2}  \nonumber\right.\\
	&& \hspace{2cm} \left. 
	+\frac{\beta}{2}\sech^2\left(\frac{\beta\omega_q }{ 2}\right) \, 
	\int_{0}^{\infty}\d\omega \, J(\omega) \, \coth\left(\frac{\beta\omega }{ 2}\right)
		\frac{\omega_q}{\omega^2-\omega_q^2}\right].
\end{eqnarray}
Apart from notational differences, \eqref{eq:sigmaX}  and \eqref{eq:sigmaZ} are the same as the Purkayastha {\it et al.} result \cite{Purkayastha2020}. 
{Expression \eqref{eq:sigmaX} also follows  from earlier work by Guarnieri \emph{et al.} \cite{Guarnieri2018} -- without the binomial approximation \eqref{eq:binomialApprox} -- who solved the Bloch-Redfield equation for the steady state of the system.}

\subsection{{Weak coupling \MFstate~for the V-system}}\label{App:Vsystem}

{Another example of the general weak coupling \MFstate~given in Eq.\ (\ref{eq:NormalizedMFGS}), is for three level atom} with two excited states $|n\rangle,n=1,2$ with energies $\omega_2=\omega_q+\tfrac{1}{2}\Delta$ and $\omega_1=\omega_q-\tfrac{1}{2}\Delta$ with $\tfrac{1}{2}\Delta\ll\omega_q$, and a ground state $|0\rangle$ with zero energy. 
The Hamiltonian of the atom is then $H_S=\omega_1|1\rangle\langle 1|+\omega_2|2\rangle\langle 2|$ and the coupling to the bath given by ${\lambda}V={\lambda}BX$ with $X=\left(|1\rangle+|2\rangle\right)\langle 0|+|0\rangle(\langle 1|+\langle 2|)$.
Since we find that $X^2=\left(|1\rangle+|2\rangle\right)\left(\langle 1|+\langle 2|\right)+2|0\rangle\langle 0|$ there will be a non-trivial energy reorganization term $X^2Q$, and we find that
\begin{equation}
	H_S'=2Q|0\rangle\langle 0|+(\omega_1+Q)|1\rangle\langle 1|+(\omega_2+Q)|2\rangle\langle 2|+Q\left(|1\rangle\langle 2|+|2\rangle\langle 1|\right).
\end{equation}

For {$H_S$ of this system we identify the energy eigenoperators}
\begin{equation}
	\begin{split}
		X_2=|2\rangle\langle 0|&\qquad \omega_2=\omega_q+\tfrac{1}{2}\Delta\\
		X_1=|1\rangle\langle 0|&\qquad \omega_1=\omega_q-\tfrac{1}{2}\Delta\\
		X_{-1}=|0\rangle\langle 1|&\qquad \omega_{-1}=-\omega_q+\tfrac{1}{2}\Delta =-\omega_1\\
		X_{-2}=|0\rangle\langle 2|&\qquad \omega_{-2}=-\omega_q-\tfrac{1}{2}\Delta =-\omega_2.
	\end{split}
\end{equation}
From the general result Eq.\ (\ref{eq:NormalizedMFGS}), we find that, since $X_mX_n^\dagger=0$ for $m$ and $n$ having opposite signs, the only pairs $(m,n)$ that will contribute to the sum {$m \neq n$} will be those for which $m$ and $n$ have the same sign, but are unequal, i.e., $(1,2),(2,1),(-2,-1),(-1,-2)$. 
{This leads to}
\begin{align}
	\rho_S^{(2)}&=\tau_S
	+\lambda^2 \omega_{12}^{-1}
	\left(\left[X_1,X_{-2} \tau_S\right]
	+\left[\tau_SX_2,X_{-1}\right]\right)
	\left(A_\beta(\omega_2)-Q\right)\notag\\
	&+\lambda^2 \omega_{21}^{-1}
	\left(\left[X_2,X_{-1} \tau_S\right]
	+\left[\tau_SX_1,X_{-2}\right]\right)
	\left(A_\beta(\omega_1)-Q\right)\notag\\
	&+\lambda^2 \omega_{-2\,-1}^{-1}
	\left(\left[X_{-2},X_1 \tau_S\right]
	+\left[\tau_SX_{-1},X_2\right]\right)
	\left(A_\beta(-\omega_1)-Q\right)\notag\\
	&+\lambda^2 \omega_{-1\,-2}^{-1}
	\left(\left[X_{-1},X_2 \tau_S\right]
	+\left[\tau_SX_{-2},X_1\right]\right)
	\left(A_\beta(-\omega_2)-Q\right)\notag\\
	&\phantom{=}+\lambda^2 
	\left[X_{-2}, \tau_SX_2\right]
	\frac{dA_\beta(\omega_2)}{d\omega_2}
	+\lambda^2\beta \tau_S\left(X_2X_{-2}
	-\tr_S\left[\tau_SX_2X_{-2}\right]\right)
	\left(A_\beta(\omega_2)-Q\right)\notag\\
	&\phantom{=}+\lambda^2 
	\left[X_{-1}, \tau_SX_1\right]
	\frac{dA_\beta(\omega_1)}{d\omega_1}
	+\lambda^2\beta \tau_S\left(X_1X_{-1}
	-\tr_S\left[\tau_SX_1X_{-1}\right]\right)
	\left(A_\beta(\omega_1)-Q\right)\notag\\
	&\phantom{=}+\lambda^2 
	\left[X_1, \tau_SX_{-1}\right]
	\frac{dA_\beta(\omega_{-1})}{d\omega_{-1}}
	+\lambda^2\beta \tau_S\left(X_{-1}X_1
	-\tr_S\left[\tau_SX_{-1}X_1\right]\right)
	\left(A_\beta(-\omega_1)-Q\right)\notag\\
	&\phantom{=}+\lambda^2 
	\left[X_2, \tau_SX_{-2}\right]
	\frac{dA_\beta(\omega_{-2})}{d\omega_{-2}}
	+\lambda^2\beta \tau_S\left(X_{-2}X_2
	-\tr_S\left[\tau_SX_{-2}X_2\right]\right)
	\left(A_\beta(\omega_{-2})-Q\right).
\end{align}
We have $\omega_{12}=-\Delta$, and $\omega_{-1\,-2}=\Delta$, and also define $\langle n|\tau_S|n\rangle=\tau_n$ and use the fact that $\tau_S$ is diagonal in the $H_S$ basis to give
\begin{align}
	\rho_S^{(2)}&=\tau_S
	+\lambda^2 \Delta^{-1}
	\left(\tau_1A_\beta(\omega_1)+\tau_0A_\beta(-\omega_1)-(\tau_2A_\beta(\omega_2)
	+\tau_0A_\beta(-\omega_2))+(\tau_1-\tau_2)Q\right)
	\left(|1\rangle\langle 2|+|2\rangle\langle 1|\right)\notag\\
	&\phantom{=}+\lambda^2 \left[\tau_0
	\frac{dA_\beta(\omega_{-2})}{d\omega_{-2}}- \tau_2
	\frac{dA_\beta(\omega_2)}{d\omega_2}\right.
	\notag\\
	&\Bigg.\phantom{=}
	+\beta\tau_2\left(A_\beta(\omega_2)-\tau_2A_\beta(\omega_2)-\tau_0A_\beta(\omega_{-2})
	-\tau_1A_\beta(\omega_1)-\tau_0A_\beta(\omega_{-1})+Q\tau_0
	\right)\Bigg]
	|2\rangle\langle 2|\notag\\
	&\phantom{=}+\lambda^2\left[\tau_0\frac{dA_\beta(\omega_{-1})}{d\omega_{-1}}
	-\tau_1\frac{dA_\beta(\omega_1)}{d\omega_1}\right.\notag\\
	&\Bigg.\phantom{=}+\beta\tau_1\left(A_\beta(\omega_1)-\tau_1A_\beta(\omega_1)
	-\tau_0A_\beta(-\omega_1)-\tau_2A_\beta(\omega_2)-\tau_0A_\beta(\omega_{-2})+Q\tau_0
	\right)\Bigg]
	|1\rangle\langle 1|\notag\\
	&\phantom{=}+\lambda^2\left[\tau_2\frac{dA_\beta(\omega_2)}{d\omega_2}
	+\tau_1\frac{dA_\beta(\omega_1)}{d\omega_1}
	-\tau_0\frac{dA_\beta(\omega_{-1})}{d\omega_{-1}}
	-\tau_0\frac{dA_\beta(\omega_{-2})}{d\omega_{-2}}\right.\notag\\
	&\Bigg.\phantom{=}-\beta\tau_0\left(\tau_2\left[A_\beta(\omega_2)
	-A_\beta(-\omega_1)-A_\beta(-\omega_{2})\right]
	+\tau_1\left[A_\beta(\omega_1)-A_\beta(-\omega_2)-A_\beta(-\omega_1)\right]\right)+Q(\tau_1+\tau_2)\Bigg]
	|0\rangle\langle 0|,
\end{align}
from which we identify the quantities $f_p(\beta)$ and $g(\beta)$ {in Eq.~\eqref{eq:Vsystem} in the main text, i.e.}
\begin{equation}	\label{eq:VweakApp}
\rho_S^{(2)}
	= \tau_S 
	+ \lambda^2 \sum_{p=0,1,2} |p \rangle  \langle p |  \, f_p(\beta) \nonumber\\
	+ \lambda^2 \left(|1\rangle\langle 2|+|2\rangle\langle 1|\right) \, g(\beta),
\end{equation}
as
\begin{align}
	f_2(\beta)&=\tau_0
	\frac{dA_\beta(\omega_{-2})}{d\omega_{-2}}-\tau_2\frac{dA_\beta(\omega_2)}{d\omega_2}
	+\beta\tau_2\big(A_\beta(\omega_2)-\tau_2A_\beta(\omega_2)-\tau_0A_\beta(\omega_{-2})
	-\tau_1A_\beta(\omega_1)-\tau_0A_\beta(\omega_{-1})+Q\tau_0\big) \nonumber \\
	f_1(\beta)&=\tau_0\frac{dA_\beta(\omega_{-1})}{d\omega_{-1}}
	-\tau_1\frac{dA_\beta(\omega_1)}{d\omega_1}+\beta\tau_1\left(A_\beta(\omega_1)-\tau_1A_\beta(\omega_1)
	-\tau_0A_\beta(-\omega_1)-\tau_2A_\beta(\omega_2)-\tau_0A_\beta(\omega_{-2})+Q\tau_0
	\right) \nonumber\\
	f_0(\beta)&=-f_1(\beta)-f_2(\beta) \nonumber\\
	g(\beta)&=\Delta^{-1}
	\left(\tau_1A_\beta(\omega_1)+\tau_0A_\beta(-\omega_1)-(\tau_2A_\beta(\omega_2)
	+\tau_0A_\beta(-\omega_2))+(\tau_1-\tau_2)Q\right).  \label{eq:gandf}
\end{align}

\subsubsection*{Low temperature limit for the V-system} 

We now evaluate the low temperature state, Eq.\ (\ref{eq:LowTempMFGS}), for the V-system.

The low temperature limit is defined by the requirement that $\beta\omega_n\gg1$ where $\omega_n$ is the energy separation between the ground state $|0\rangle$ and any excited state of the system, in which case the {thermal} state $\tau_S$ is {approximated as}  $|0\rangle\langle 0|$. For the V-system, given $\Delta\ll\omega_q$, this translates into the condition $\beta\omega_q\gg1$.  
In addition, in arriving at the general expression {\eqref{eq:VweakApp}} for the V-system density operator, limits are imposed on $\beta$ in order to satisfy the conditions for the binomial approximation, see Eq.~\eqref{eq:binomialApproxCondition}. At low temperatures, this condition reduces to
{
\begin{equation}
	\left|\beta\lambda^2\sum_{n}\langle 0|X_nX_n^\dagger|0\rangle\left(A_\beta(\omega_n)_{\beta\to\infty}-Q\right)\ll1  \right|
\end{equation}
which for the V-system becomes, with $\beta=1/T$,
\begin{equation}\label{eq:lowestLowTemp}
	T\gg \left| 2\lambda^2\left(\int_{0}^{\infty}\d\omega \, J(\omega)\frac{\omega+\omega_q}{(\omega+\omega_q)^2-\tfrac{1}{4}\Delta^2}-Q\right) \right|.
\end{equation}
Thus at low temperatures for which $\beta\omega_q\gg1$ that also satisfy this condition \eqref{eq:lowestLowTemp}, the \MFstate~} is then, from Eq.\ (\ref{eq:LowTempMFGS}), 
\begin{equation}
	\begin{split}
		\lowT=&|0\rangle\langle 0|
		\left(1-\int_{0}^{\infty} \d\omega \, 2\lambda^2J(\omega)\frac{\left(\omega+\omega_q\right)^2+\frac{1}{4}\Delta^2}{((\omega+\omega_q)^2-\frac{1}{4}\Delta^2)^2}\right)\\
		&+|1\rangle\langle 1|\int_{0}^{\infty}\d\omega \, \frac{\lambda^2J(\omega)}{(\omega+\omega_q-\frac{1}{2}\Delta)^2}
		+|2\rangle\langle 2|\int_{0}^{\infty}\d\omega \, \frac{\lambda^2J(\omega)}{(\omega+\omega_q+\frac{1}{2}\Delta)^2}\\
		&+\left(|1\rangle\langle 2|+|2\rangle\langle 1|\right)\int_{0}^{\infty}\d\omega \, \frac{\lambda^2J(\omega)}{(\omega+\omega_q)^2-\frac{1}{4}\Delta^2}.
	\end{split}
\end{equation}
Of significance is the presence of coherence between the pair of upper levels, induced, in this case, by the {quantum fluctuations that are dominant at low temperatures}.

\section{Ultrastrong coupling limit}\label{app:StrongCouplingDetails}

The approach taken here is based on a straightforward idea that allows the description of the quantum Zeno effect, historically first analysed for a system subject to rapidly repeated measurement\cite{Misra1977}, but later shown to be equivalent to a strong coupling model \cite{Facchi2003a}. The method is built around standard perturbation theory \cite{Facchi2003}, but can be put on a more formal basis including the estimation of errors \cite{Burgarth2018}.

The idea is to find an approximate form for the time evolution operator $U_\lambda(t)=\exp(-itH_\lambda )$ with  total Hamiltonian $H_\lambda=H_S+\lambda V$, where $H_S$ is strongly perturbed by the term $\lambda V$, where $\lambda$ is a parameter that scales the strength of the perturbation. The desired limit is the strong perturbation limit,
\begin{equation}
	U_\infty(t)=\lim_{\lambda\to\infty}e^{-it(H_S+\lambda V)}.
\end{equation}
The analysis of the case of unitary dynamics is outlined in the following Section for the purposes of illustrating the approach used. Further developments needed to deal with the strong coupling limit of the \MFstate~are presented in Section \ref{app:StrongCouplingGibbsStateDetails}. Specifically, we will transition to the thermal equilibrium case via the `Wick rotation', $t\to-i\beta$, as well as a trace over the reservoir states.

\subsection{Strong coupling for unitary dynamics}\label{app:Zeno}
 
The idea is to write the evolution operator as $U_\lambda(t)=\exp\left[-it\lambda \left(V+\lambda^{-1}H_S\right)\right]$ and work with the eigenstates and eigenvalues of the Hamiltonian $\tilde{H}_\lambda=V+\lambda^{-1}H_S$ where now $V$ is the `unperturbed system' and $\lambda^{-1}H_S$ is the `perturbation', an appropriate description for $\lambda$ large.

The procedure then is to identify the `unperturbed' eigenstates of $V$, $|v_n\rangle$ say, and corresponding eigenvalues $v_n$ (ignoring here any considerations concerning degeneracies), from which by first order perturbation theory the perturbed eigenvalues
\begin{equation}\label{eq:perturbedeigenvalues}
	\tilde{v}_n=v_n+\lambda^{-1}\langle v_n|H_S|v_n\rangle+\mathcal{O}(\lambda^{-2})
\end{equation}
are determined. The eigenstates $|\tilde{v}_n\rangle$ of $V+\lambda^{-1}H_S$ can also be calculated to lowest order, but are not needed for determining the leading contribution to $U_\infty(t)$, except to note that $|\tilde{v}_n\rangle=|v_n\rangle+\mathcal{O}(\lambda^{-1})$. We then write $U_\lambda(t)=\sum_{n}U_\lambda(t)\tilde{P}_n$ where $\tilde{P}_n=|\tilde{v}_n\rangle\langle \tilde{v}_n|$. Further
\begin{equation}
	\sum_{n}U_\lambda(t)\tilde{P}_n=\sum_{n}\exp\left(-i\lambda t\tilde{H}_\lambda\right)\tilde{P}_n
	=\sum_{n}\exp\left(-it\lambda \tilde{v}_n\right)\tilde{P}_n.
\end{equation}
Substituting for $\tilde{v}_n$ from Eq.\ (\ref{eq:perturbedeigenvalues}), and noting that $\tilde{P}_n=P_n+\mathcal{O}(\lambda^{-1})$ where $P_n=|v_n\rangle\langle v_n|$, we find, after a little manipulation
\begin{equation}\label{eq:manipulation}
	U_\lambda(t)=\sum_{n}\exp\left[-it\left(\langle v_n|H_S|v_n\rangle+\lambda v_n+\mathcal{O}(\lambda^{-1})\right)\right]P_n+\mathcal{O}(\lambda^{-1}).
\end{equation}
There are expressions available for the $\mathcal{O}(\lambda^{-1})$ corrections, but they are unfeasibly complicated to evaluate, so the results reported here will be for the ultrastrong `$\lambda\to\infty$' limit, for which we can write
\begin{equation}
	\lim_{\lambda\to\infty}U_\lambda(t) \, e^{it\lambda V}=\exp\left[-it\sum_{n}P_nH_SP_n\right]
\end{equation}
Thus the effect of the strong coupling is to replace the original system Hamiltonian $H_S$ by a `partitioned' Hamiltonian, i.e.,
\begin{equation}\label{eq:partitionedHamiltonian}
	H_S\to\sum_{n}P_nH_SP_n.
\end{equation}.

{We note that for a system with Hamiltonian $H_S$ (and no coupling to an environment)  following unitary evolution with $H_S$ but experiencing perturbations $\epsilon V_S$ with $||V_S||=1$, some of the authors of the unitary Zeno effect techniques  \cite{Facchi2003a,Facchi2003, Burgarth2018} have recently shown that the unique robust stationary state at long times is the Gibbs state  \cite{Burgarth2020a}, i.e.,
\begin{equation}
	e^{- i t (H_S + \epsilon V_S)} \, e^{-\beta H_S} \, e^{+ i t (H_S + \epsilon V_S)} 
	\sim e^{-\beta H_S} + {\cal O} (\epsilon).
\end{equation}
}

\subsection{Mean force \Gib~state at ultrastrong coupling to a reservoir} \label{app:StrongCouplingGibbsStateDetails}

The aim here is to evaluate the \MFstate~in the strong coupling limit, i.e., determining the large $\lambda$ limit of
\begin{equation}
	\tilde{\rho}_{S}=\tr_R\left[e^{-\beta \left(H_S+H_R+\lambda BX+\lambda^2X^2Q\right)}\right],
\end{equation}
where as before, $H_R$ is the Hamiltonian of the reservoir, and $V=\lambda BX$ is the coupling between a reservoir operator $B$ and a system operator  $X$. In general, the reservoir will have a quasi-continuous energy spectrum, necessary for the system-reservoir to evolve to a steady state. The interaction $XB$ must be such that energy is exchanged between the system and the reservoir so that an energetic equilibrium state can be established. As before, the reservoir is assumed to be bosonic, with $B$ of the form
\begin{equation}
	B=\int_{0}^{\infty}\d\omega\sqrt{J(\omega)}\left(b(\omega)+b^\dagger(\omega)\right),
\end{equation}
where $\left[b(\omega),b^\dagger(\omega')\right]=\delta(\omega-\omega')$ and where $J(\omega)$ is the spectral density.

As well as the replacement $t\to-i\beta$ as compared to the unitary case, an extra step required here is to carry out the trace over the reservoir states.
The starting point is the full expression for the unnormalized joint system-reservoir \Gib~state
\begin{equation}
	\tilde{\rho}_{SR}=e^{-\beta(H_S+H_R+\lambda BX+\lambda^2X^2Q)}.
\end{equation}We first separate out the reservoir Hamiltonian by moving to an interaction picture:
\begin{equation} \label{eq:rhoSR}
	\tilde{\rho}_{SR}=e^{-\beta H_R}W(\beta),
\end{equation}
where
\begin{equation}
	W(\beta)=e^{\beta H_R}e^{-\beta(H_S+H_R+\lambda BX+\lambda^2X^2Q)}.
\end{equation}
Differentiating with respect to $\beta$ then gives
\begin{equation} \label{eq:diff}
	W'(\beta)=-(H_S+\lambda X B(-i\beta)+\lambda^2X^2Q)W(\beta),
\end{equation}
where we have introduced the `interaction picture operator'
\begin{equation}
	B(-i\beta)
	=e^{\beta H_R}B e^{-\beta H_R}
	=\int_{0}^{\infty}\d\omega \sqrt{J(\omega)}\left(b(\omega)e^{-\beta\omega}+b^\dagger(\omega)e^{\beta\omega}\right).
\end{equation}
Eq.~\eqref{eq:diff} has as its solution a `time ordered' or better, a `beta ordered', exponential, with $W(0)=1$
\begin{equation}\label{eq:betaOrderedExpo}
	W(\beta)=\mathbb{T}\exp\left[-\int_{0}^{\beta}\d\beta'(H_S+\lambda X B(-i\beta')+\lambda^2X^2Q)\right]
	\equiv \mathbb{T}\exp\left[-\int_{0}^{\beta}\d\beta'  \lambda^{-1} \tilde{H}_\lambda(\beta') \right].
\end{equation}
Here we have defined the Hamiltonian
\begin{equation}\label{eq:lambdaHam}
	\tilde{H}_\lambda(\beta)=V_\lambda+\lambda^{-1}H_S
\end{equation}
with $V_\lambda=B(-i\beta)X+\lambda X^2Q$ being the `unperturbed Hamiltonian' and $\lambda^{-1}H_S$ the `perturbation'.
We will now work with the eigenstates and eigenvalues of $\tilde{H}_\lambda(\beta)$. 

Note that $B(-i\beta)$ commutes with all the system operators appearing in the above expression for $\tilde{H}_\lambda(\beta)$, and hence will be treated as a c-number parameter in what follows. Thus, the eigenstates of the unperturbed Hamiltonian $V_\lambda$ will be simply the eigenstates of $X$, that is $|x_n\rangle$, with eigenvalue $v_\lambda$, i.e.,
\begin{equation}
	V_\lambda|x_n\rangle=v_\lambda|x_n\rangle=\left(B(-i\beta)x_n+\lambda x_n^2Q\right)|x_n\rangle.
\end{equation}
Substituting \eqref{eq:lambdaHam} into \eqref{eq:betaOrderedExpo} and expanding we have
\begin{equation}
	W(\beta)=1+\sum_{j=1}^{\infty}\int_{0}^{\beta}\d\beta_j\ldots \int_{0}^{\beta_2}\d\beta_1 \, (-\lambda)^j\tilde{H}_\lambda(\beta_j)\ldots \tilde{H}_\lambda(\beta_1).
\end{equation}
If we let $|\tilde{v}_n(\beta)\rangle$ be the eigenstates of $\tilde{H}_\lambda(\beta)$ with eigenvalues $\tilde{v}_n(\beta)$, and define $\tilde{P}_n(\beta)$ as the projection operator onto these eigenstates
\begin{equation}
	\tilde{P}_n(\beta)=|\tilde{v}_n(\beta)\rangle\langle \tilde{v}_n(\beta)|,
\end{equation}
then we can write
\begin{equation}
	\tilde{H}_\lambda(\beta)=\sum_{n}\tilde{v}_{n}(\beta)\tilde{P}_n(\beta).
\end{equation}
The expansion of $W(\beta)$ then becomes
\begin{equation}\label{eq:expandedW}
	W(\beta)=1+\sum_{j=1}^{\infty}\int_{0}^{\beta}\d\beta_j\ldots \int_{0}^{\beta_2}\d\beta_1 \, \left(-\lambda\right)^j
	\sum_{n_j}\ldots \sum_{n_1}\tilde{v}_{n_j}(\beta_j)\ldots \tilde{v}_{n_1}(\beta_1)\tilde{P}_{n_j}(\beta_j)\ldots \tilde{P}_{n_1}(\beta_1).
\end{equation}
By first order perturbation theory, given that the unperturbed eigenstates are the eigenstates of $X$, and the perturbation $\lambda^{-1}H_S$ is $\mathcal{O}(\lambda^{-1})$, it follows that $|\tilde{v}_n(\beta)\rangle=|x_n\rangle+\mathcal{O}(\lambda^{-1})$. Hence $\tilde{P}_n(\beta)=P_n+\mathcal{O}(\lambda^{-1})$
where $P_n=|x_n\rangle\langle x_n|$ are the projectors onto the eigenstates of $X$, so that the product of projection operators appearing in \eqref{eq:expandedW} will become
\begin{equation}
	\tilde{P}_{n_j}(\beta_j)\ldots \tilde{P}_{n_1}(\beta_1)=P_{n_j}\ldots P_{n_1}+\mathcal{O}(\lambda^{-1}).
\end{equation}
As the eigenstates of $X$ will be orthornormal, the product of projection operators will collapse to a single operator, so the sum over $(n_j\ldots n_1)$ in \eqref{eq:expandedW} will reduce to
\begin{equation}
	W(\beta)=1+\sum_{n}\sum_{j=1}^{\infty}\int_{0}^{\beta}\d\beta_j\ldots \int_{0}^{\beta_2}\d\beta_1 \, \left(-\lambda\right)^j
	\tilde{v}_{n}(\beta_j)\ldots \tilde{v}_{n}(\beta_1)P_{n}+\mathcal{O}(\lambda^{-1}).
\end{equation}
Further, by first order perturbation theory, the eigenvalue $\tilde{v}_n(\beta)$ of $\tilde{H}_\lambda$ can be expressed in terms of the eigenvalue $v_n(\beta)=x_nB(-i\beta)+\lambda x_n^2Q$ of the unperturbed Hamiltonian, plus a first order correction i.e.,
\begin{equation}
	\tilde{v}_n(\beta)=x_nB(-i\beta)+\lambda x_n^2Q+\lambda^{-1}\langle x_n|H_S|x_n\rangle+\mathcal{O}(\lambda^{-2}),
\end{equation}
and we now have
\begin{align}
	W(\beta)&=1+\sum_{n}\sum_{j=1}^{\infty}\int_{0}^{\beta}\d\beta_j\ldots \int_{0}^{\beta_2}\d\beta_1
	\left(\langle x_n|H_S|x_n\rangle+\lambda x_nB(-i\beta_j)+\lambda^2 x_n^2Q+\mathcal{O}(\lambda^{-1})\right)P_n\ldots \notag\\
	&\hspace*{4cm}\times\left(\langle x_n|H_S|x_n\rangle+\lambda x_nB(-i\beta_1)+\lambda^2 x_n^2Q+\mathcal{O}(\lambda^{-1})\right)P_{n}+\mathcal{O}(\lambda^{-1})\notag\\
	&=1+\sum_{n}\sum_{j=1}^{\infty}\int_{0}^{\beta}\d\beta_j\ldots \int_{0}^{\beta_2}\d\beta_1
	\left(P_nH_SP_n+\lambda XB(-i\beta_j)+\lambda^2 X^2Q+\mathcal{O}(\lambda^{-1})\right)P_n\ldots \notag\\
	&\hspace*{4cm}\times\left(P_nH_SP_n+\lambda XB(-i\beta_1)+\lambda^2 X^2Q+\mathcal{O}(\lambda^{-1})\right)P_{n}+\mathcal{O}(\lambda^{-1})\notag\\
	&=\mathbb{T}\exp\left[-\int_{0}^{\beta}\d\beta'\left(\sum_{n}P_nH_SP_n+XB(-i\beta')+\lambda^2X^2Q+\mathcal{O}(\lambda^{-1})\right)\right]+\mathcal{O}(\lambda^{-1}).
\end{align}
If we now assume $\lambda$ is sufficiently large that terms $\mathcal{O}(\lambda^{-1})$ can be ignored, then we can write for $\tilde{\rho}_{SR}$ from Eq.~\eqref{eq:rhoSR}
\begin{eqnarray}\label{eq:strongLimit}
	\tilde{\rho}_{SR}
	&=&e^{-\beta H_R} \,\, \mathbb{T}\exp\left[-\int_{0}^{\beta}\d\beta'\left(\sum_{n}P_nH_SP_n+XB(-i\beta')+\lambda^2X^2Q\right)\right] \nonumber\\
	&=& e^{-\beta H_R} \,\, \mathbb{T}\exp\left[- \int_{0}^{\beta}\d\beta' \, e^{\beta' H_R}\left(\sum_{n}P_nH_SP_n+XB+\lambda^2X^2Q\right) e^{-\beta' H_R} \right] \\
	&=&\exp\left[-\beta\left(\sum_{n}P_nH_SP_n+H_R+\lambda XB+\lambda^2X^2Q\right)\right].
\end{eqnarray}

As the pair of exponents $H_R+\lambda XB$ and $P_n \left(H_S+\lambda^2X^2Q\right)P_n$ in \eqref{eq:strongLimit} commute, we can write
\begin{equation}\label{eq:StrongCouplingGibbs}
	\tilde{\rho}_{SR}=\exp\left[-\beta\left(H_R+\lambda XB\right)\right] \, \exp\left(-\beta\sum_{n}P_n \left(H_S+\lambda^2X^2Q\right)P_n\right),
\end{equation}
which now makes it possible to carry out the trace over the reservoir states, i.e.,
\begin{equation}\label{eq:MFGSStart}
	\tilde{\rho}_S=\tr_R\left[\exp\left(-\beta\left(H_R+\lambda XB\right)\right)\right] \, \exp\left(-\beta\sum_{n}P_n \left(H_S+\lambda^2X^2Q\right)P_n\right).
\end{equation}

The first factor here involving the trace can be evaluated directly by rewriting, in the exponent, $H_R+\lambda XB=H_R+\lambda XB+\lambda^2X^2Q-\lambda^2X^2Q=H_{SR}-H_S-\lambda^2X^2Q$ where $H_{SR}$ is the total Hamiltonian of the combined system and reservoir, equation \eqref{eq:totalHamiltonian}, and $Q=\int_{0}^{\infty}d\omega J(\omega)/\omega $ is the reorganization energy. From equation \eqref{eq:totalHamiltonian} we can read off the expression for the difference $H_{SR}-H_S$ so that this trace can then be written
\begin{align}
	\text{Tr}_R\left[\exp\left({-\beta\left(H_R+\lambda XB\right)}\right)\right]
	&= \text{Tr}\left[
	\exp\left(
	-\frac{\beta}{2} 
	\int_{0}^\infty \d\omega \, 
	\left(p^2(\omega)+\left(\omega q(\omega)+\lambda \sqrt{\frac{2J(\omega)}{\omega}} X \right)^2\right) 
	\right)
	\right]  
	\, e^{\beta\lambda^2X^2Q}\nonumber\\
	&=Z_R \, e^{\beta\lambda^2X^2Q}.
\end{align}
{For large $\lambda$ of course this expression arising from taking the trace over the reservoir diverges. However, when multiplying it with the other $\lambda$-dependent term in \eqref{eq:MFGSStart}, the divergent factor is exactly cancelled since $X^2 = \sum_n P_n X^2 P_n$. This leaves a finite expression already before normalisation of the state, i.e.}
the unnormalised system state is, from Eq.\ (\ref{eq:MFGSStart}) and dropping the constant $Z_R$,
\begin{equation}
	\tilde{\rho}_S
	= e^{\beta\lambda^2X^2Q} \, e^{-\beta \sum_{n}P_n \left(H_S+\lambda^2X^2Q\right)P_n} 
	{= e^{-\beta \sum_{n}P_n \, H_S \, P_n}}.
\end{equation}
{Thus all} dependence on $\lambda$ is removed, and we get the strong coupling limit for the normalised \MFstate~
\begin{equation} \label{eq:ultrastrong}
	\rho_S =\frac{e^{-\beta\sum_{n}P_n H_S P_n} }{ \tr\left[e^{-\beta\sum_{n}P_n H_S P_n}\right]}.
\end{equation}
This is result \eqref{eq:PartitionedState} in the main text.

\medskip
\medskip

Our derived MFG state can be compared to the steady state \emph{conjectured}  by Goyal and Kawai \cite{Goyal2019}. They propose that in the strong coupling limit a system relaxes to a stationary state  $\lim_{\lambda\to\infty} \stst_S (t \to \infty)$ given by
\begin{equation} \label{eq:GKstate}
	\varrho_S^{conj}=\sum_{n}P_n \, \tau_S(\beta) \, P_n,
\end{equation}
where $\tau_S(\beta)$ is the \Gib~state with respect to the bare system Hamiltonian $H_S$. Note that for both states,  \eqref{eq:ultrastrong} and \eqref{eq:GKstate}, the projection operators $P_n$ appear. 
For comparison, equation \eqref{eq:ultrastrong} can be rewritten as
\begin{equation}
	\rho_S \propto e^{-\beta\sum_{n}P_nH_SP_n}=\sum_{n}^{}P_n \, e^{-\beta\langle x_n|H_S|x_n\rangle}
\end{equation}
which gives a weighting of $e^{-\beta\langle x_n|H_S|x_n\rangle}$ accorded to each of the subspaces defined by the projection operator $P_n$. This can be compared to steady state conjecture \eqref{eq:GKstate}
\begin{equation}
	\varrho_S^{conj} \propto \sum_{n}^{}P_n e^{-\beta H_S}P_n=\sum_{n}^{}P_n \, \langle x_n|e^{-\beta H_S}|x_n\rangle.
\end{equation}
While similar, there are clearly some differences. The interpretation of the derived ultrastrong coupling expression \eqref{eq:ultrastrong}  is that the equilibrium state is a Gibbs state w.r.t. a modified Hamiltonian, $\sum_{n}P_nH_SP_n$. The effect of the ultrastrong coupling is to force the system to equilibrate according to the eigenstates of the now dominant system interaction operator $X$, rather than with respect to the system Hamiltonian, with a weighting according to the mean value of the system Hamiltonian with respect to each such eigenstate.
 
Below the two states will be compared for two specific situations, that of a single qubit strongly coupled to a single reservoir Section \ref{sub:singlequbit}, and the more general case of two coupled qubits which are individually coupled to two independent reservoirs, Section \ref{sub:twoqubits}.

\subsection{Two systems and two reservoirs} \label{sub:twoSystemsTwoReservoirs}

This generalization of the previous case can take two forms, one in which a single system is coupled to more than one reservoir, {such as a spin system coupled to a three dimensional reservoir \cite{Anders2020}}, the other in which separate systems (that may interact with each other)  each  individually couple to a separate reservoir. Within these two possibilities there is a further generalization to reservoirs held at different temperatures. This last case would lead to non-equilibrium steady states, a different class of problem, and will not be considered here. The first case will also not be considered as it has been found to introduce difficulties that also require further development of the theory. Thus here we will focus on the case of two  interacting systems, $S_\alpha$ with $\alpha=1,2$, each individually interacting with a separate reservoir, $R_\alpha$, and these reservoirs are at the same inverse temperature $\beta$.

In this case, we will have a Hamiltonian of the form
\begin{equation}
	H_{\lambda_1\lambda_2}=H_{S_1}+H_{S_2}+H^{int}_{12}+H_{R_1}+H_{R_2}+\lambda_1 B_1X_1+\lambda_2 B_2X_2+\lambda_1^2X_1^2Q_1+\lambda_2^2X_2^2Q_2,
\end{equation}
where the interaction between the systems is given by $H^{int}_{12}$, and the two reservoirs are independent, so that $\left[B_1,B_2\right]=0$. The system Hamiltonian is now $H_S=H_{S_1}+H_{S_2}+H^{int}_{12}$, and the $Q_\alpha$ are the reorganization energies for the coupling of each system to its respective reservoir, i.e.,
\begin{equation}
	Q_\alpha=\int_{0}^{\infty}\d\omega \, \frac{J_\alpha(\omega)}{\omega},
\end{equation}
where $J_n(\omega)$ is the spectral density of the coupling between $S_\alpha$ and $B_\alpha$.

Taking the global thermal state of systems and reservoirs at inverse temperature $\beta$, the \MFstate~of the combined system $S\equiv S_1\oplus S_2$ is formally,
\begin{equation}
	\rho_S=\frac{\tr_{B_1B_2}
	\left[e^{-\beta H_{\lambda_1\lambda_2}}\right]}{\tr _{S_1S_2}\left[ \tr_{B_1B_2}
	\left[e^{-\beta H_{\lambda_1\lambda_2}}\right] \right] }.
\end{equation}
The strong coupling limit can now be performed as in the single reservoir case, but sequentially for the two couplings and in either order. E.g., first with respect to the parameter $\lambda_1$, giving rise to a partitioning of the Hamiltonian in terms of the projection operators $P_{1m}$ onto the eigenstates of $X_1$, and then repeated with respect to the parameter $\lambda_2$, and the concomitant projection operators onto the eigenstates of  $X_2$. Note that $\left[P_{1m},P_{2n}\right]=0$ as the systems $S_1$ and $S_2$ are independent.

The traces over the two independent reservoir states can each be done as in the single reservoir case. The result is that the unnormalized joint state for the combined system $S$ is given by
\begin{equation} \label{eq:twobathsapp}
	\tilde{\rho}_S=\exp\left[-\beta\sum_{mn}P_{1m}P_{2n}H_SP_{2n}P_{1m}\right],
\end{equation}
where $P_{\alpha n}=|x_{\alpha n}\rangle\langle x_{\alpha n}|$, and $X_{\alpha}|x_{\alpha m}\rangle=x_{\alpha m}|x_{\alpha m}\rangle$, $\alpha=1,2$. This is the unnormalised version of result Eq.~\eqref{eq:twobaths} in the main text.

To see the different contributions of the bare Hamiltonians $H_{S\alpha}$ and the inter-system coupling $H^{int}_{12}$, one may further expand
\begin{equation}
	\sum_{mn}P_{1m}P_{2n}H_{S}P_{2n}P_{1m}=\sum_{\alpha,m}P_{\alpha m}H_{S_\alpha}P_{\alpha m}+\sum_{mn}P_{1m}P_{2n}H^{int}_{12}P_{2n}P_{1m},
\end{equation}
and as the various terms in this expression all mutually commute, we have
\begin{equation}\label{eq:TwoSystemsTwoReservoirs}
	\tilde{\rho}_S=\exp\left[-\beta\sum_{\alpha,m}P_{\alpha m}H_{S_\alpha}P_{\alpha m}\right]\exp\left[-\beta\sum_{mn}P_{1m}P_{2n}H^{int}_{12}P_{2n}P_{1m}\right].
\end{equation}

Interestingly, the structure of $\tilde{\rho}_S$ in \eqref{eq:twobathsapp} is exactly the same as that for the single reservoir case, cf. \eqref{eq:PartitionedState}, but with the single reservoir projectors $P_n$ replaced by  the tensor product $P_{1m} \, P_{2n}$.
It can be noted that for two coupled systems interacting with a common reservoir, as studied in, for instance \cite{Orth2010,Benatti2010,Deng2016} the same result, \eqref{eq:TwoSystemsTwoReservoirs} will follow.

\section{Applications of strong coupling result} \label{App:strongapps}

The strong coupling results \eqref{eq:ultrastrong} and \eqref{eq:twobathsapp} are applied in three scenarios, a single qubit coupled to a single reservoir, a V-system coupled to a single reservoir, and a pair of interacting qubits each coupled to two independent reservoirs at the same temperature.

\subsection{Single qubit} \label{sub:singlequbit}

\begin{figure}[b]
\includegraphics[trim=1.15cm 0.1cm 0cm 0cm,clip, width=0.6\textwidth]{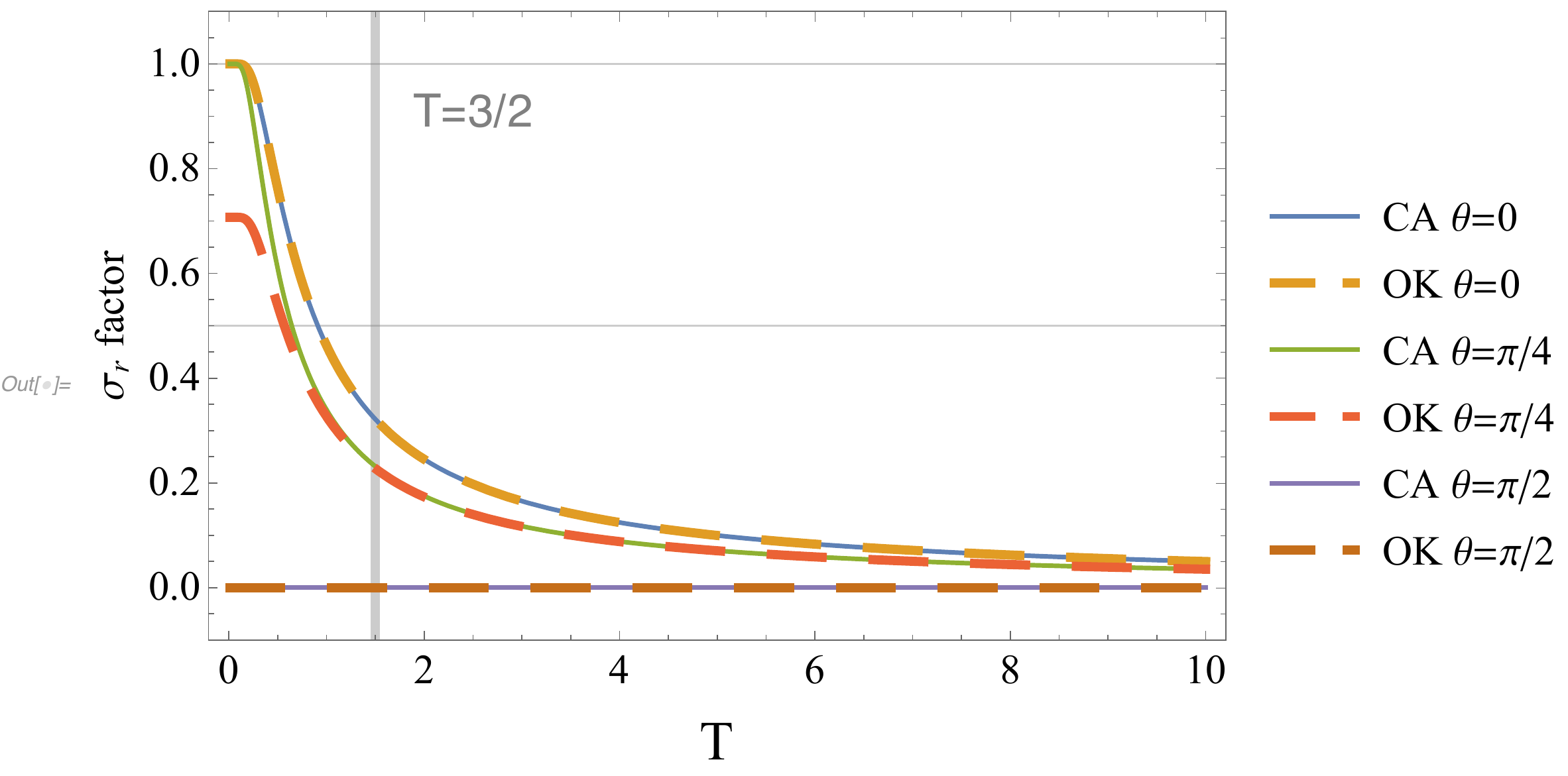}
\caption{\label{fig:singlequbit} 
$\sigmar$-prefactor $\tanh\left(\tfrac{1}{2}\beta\omega_q\cos\theta\right)$ in \eqref{eq:CAsingle} (solid lines) and the corresponding prefactor $\cos\theta \tanh\left(\tfrac{1}{2}\beta\omega_q\right)$ in \eqref{eq:OK} \cite{Orman2020} (dashed lines) are plotted for three angles $\theta=0, \pi/4, \pi/2$ as functions of  temperature $T=1/\beta$, for $\omega_q=1$. While for  large temperatures  these two expressions are approximately the same, for low temperatures  they differ noticeably for angles $0 < \theta <  \frac{\pi }{ 2}$. For $\lambda=5$, Orman and Kawai (OK) have fully solved the spin dynamics numerically, using the method of hierarchical equations of motion (HEOM). At  temperature value $T=3/2$ (grey), and for the above choices of angles $\theta$, they showed that the dynamical steady state $\stst_S (t \to \infty)$ is numerically close to \eqref{eq:OK}. As can be seen,  differences between \eqref{eq:OK} and \eqref{eq:CAsingle} would not be distinguishable for this $T$-value. }
\end{figure}

As an example for \eqref{eq:ultrastrong}, we again consider the single qubit Hamiltonian $H_S=\tfrac{1}{2}\omega_q\sigma_z$ and the system operator coupling to the single reservoir is given by $X=\boldsymbol{\sigma}\cdot\hat{\mathbf{r}}$ where $\hat{r}$ is a unit vector $\hat{r}=\sin\theta\cos\phi \, \hat{x}+\sin\theta\sin\phi \, \hat{y}+\cos\theta \, \hat{z}$ (cf. \App~\ref{App:spinboson} and \cite{Purkayastha2020}). The eigenstates of $X$ will be the states $|\pm_{\hat{\mathbf{r}}}\rangle$, with $\boldsymbol{\sigma}\cdot\hat{\mathbf{r}}|\pm_{\hat{\mathbf{r}}}\rangle=\pm|\pm_{\hat{\mathbf{r}}}\rangle$, and the required projection operators are $P_\pm=|\pm_{\hat{\mathbf{r}}}\rangle\langle\pm_{\hat{\mathbf{r}}}|$. We find that
\begin{equation}
	\sum_{n=\pm}^{}P_nH_SP_n= \sigmar \, \tfrac{1}{2}\omega_q\cos\theta, \end{equation}
and a straightforward calculation then gives,
\begin{equation} \label{eq:CAsingle}
	\rho_S=\tfrac{1}{2}\left(1- \sigmar  \, \tanh\left(\tfrac{1}{2}\beta\omega_q\cos\theta\right)\right).
\end{equation}
This is equation \eqref{eq:singlequbit} in the main text, and Fig.~\ref{fig:singlequbit}  shows plots of the $\sigmar$-prefactor of \eqref{eq:CAsingle} (solid) as a function of $T$. 

For comparison, the conjectured state \eqref{eq:GKstate} evaluated by Orman and Kawai (OK) \cite{Orman2020} for this system is 
\begin{equation}  \label{eq:OK}
	\varrho_S^{OK}=\tfrac{1}{2}\left(1-  \sigmar \, \cos\theta\tanh\left(\tfrac{1}{2}\beta\omega_q\right)\right), 
\end{equation}
which differs from \eqref{eq:CAsingle} only in the positioning of $\cos \theta$. 
Plots of the $\sigmar$-prefactors of \eqref{eq:CAsingle} and \eqref{eq:OK} are shown in Fig.~\ref{fig:singlequbit}, together with the numerical temperature at which OK confirmed steady state convergence to the state \eqref{eq:OK}. One can see that at the temperature they tested with HEOM, the dynamical convergence of the system state to \eqref{eq:OK}, for three angles $\theta=0, \pi/4, \pi/2$, could  equally be convergence to \eqref{eq:CAsingle} instead.

\subsection{V-system} \label{sub:strongCVsystem}

For the three-level V-system,  see Appendix \ref{App:Vsystem}, the system's bath-coupling operator $X$ has eigenstates
\begin{equation}
		|x_0\rangle=\frac{1}{\sqrt{2}}\left(|1\rangle-|2\rangle\right),\quad|x_\pm\rangle
		=\frac{1}{2}\left(\pm\sqrt{2}|0\rangle+|1\rangle+|2\rangle\right)	,
\end{equation}
from which we can construct the relevant projection operators $P_0=|x_0\rangle\langle x_0|$ and $P_\pm=|x_\pm\rangle\langle x_\pm|$. We find that, with $H_S=\omega_2|2\rangle\langle 2|+\omega_1|1\rangle\langle 1|$ and $\omega_2=\omega_q+\tfrac{1}{2}\Delta,\omega_1=\omega_q-\tfrac{1}{2}\Delta$
\begin{equation}
		\sum_{n}P_n H_S P_n=\tfrac{1}{2}\omega_q\left( \mathbbm{1}+|x_0\rangle\langle x_0|\right),
\end{equation}
from which it follows that
\begin{equation} \label{eq:Vstrong}
	\begin{split}
		\rho_S=\left(2+e^{-\frac{1}{2}\beta\omega_q}\right)^{-1}\Big[|0\rangle\langle 0|\Big.&+\tfrac{1}{2}\left(1+e^{-\frac{1}{2}\beta\omega_q}\right)\left(|1\rangle\langle 1|+|2\rangle\langle 2|\right)\\
		&+\left.\tfrac{1}{2}\left(1-e^{-\frac{1}{2}\beta\omega_q}\right)\left(|1\rangle\langle 2|+|2\rangle\langle 1|\right)\right].
	\end{split}
\end{equation}
Thus,  in the strong coupling limit coherence persists between the pair of excited states $|1\rangle$ and $|2\rangle$, which is largest at low temperature $\beta\omega_q\gg1$, and vanishes in the limit of high temperatures, $\beta\omega_q\ll1$.

\subsection{Two interacting qubits, two reservoirs}\label{sub:twoqubits}

Here we will apply the strong coupling result \eqref{eq:twobathsapp} to the example of two identical qubits that are coupled to each other, and individually coupled to separate reservoirs at the same temperature.
The system Hamiltonian is given by
\begin{equation}
	H_S=\tfrac{1}{2}\omega_q\left(\sigma_{1z}+\sigma_{2z}\right)+\lambda_S\left(\sigma_{1+}\sigma_{2-}+\sigma_{1-}\sigma_{2+}\right),
\end{equation}
where for qubit $\alpha=1,2$, the operators are $\sigma_{\alpha z}=|e_\alpha\rangle\langle e_\alpha|-|g_\alpha\rangle\langle g_\alpha|$ and $H^{int}_{12}=\lambda_S\left(\sigma_{1+}\sigma_{2-}+\sigma_{1-}\sigma_{2+}\right)$ is the qubit interaction term with $\lambda_S$ the inter-qubit coupling of arbitrary strength and  $\sigma_{\alpha -}=|g_\alpha\rangle\langle e_\alpha|=\sigma_{\alpha +}^\dagger$.
The coupling of the qubits to the two reservoirs is given by
\begin{equation}
	\lambda_1B_1X_1+\lambda_2B_2X_2,
\end{equation}
with $X_\alpha=\sigma_{\alpha x}=\sigma_{\alpha +}+\sigma_{\alpha -}$ and $\lambda_\alpha$ both large coupling constants. The normalized eigenstates of $X_\alpha$ are $|\pm_\alpha\rangle=\left(|e_\alpha\rangle\pm|g_\alpha\rangle\right)/\sqrt{2}$ in terms of which can be defined the projection operators $P_{\alpha \pm}=|\pm_\alpha\rangle\langle \pm_\alpha|$. 

\begin{figure}[t]
\includegraphics[trim=1.15cm 0.1cm 0cm 0cm,clip, width=0.65\textwidth]{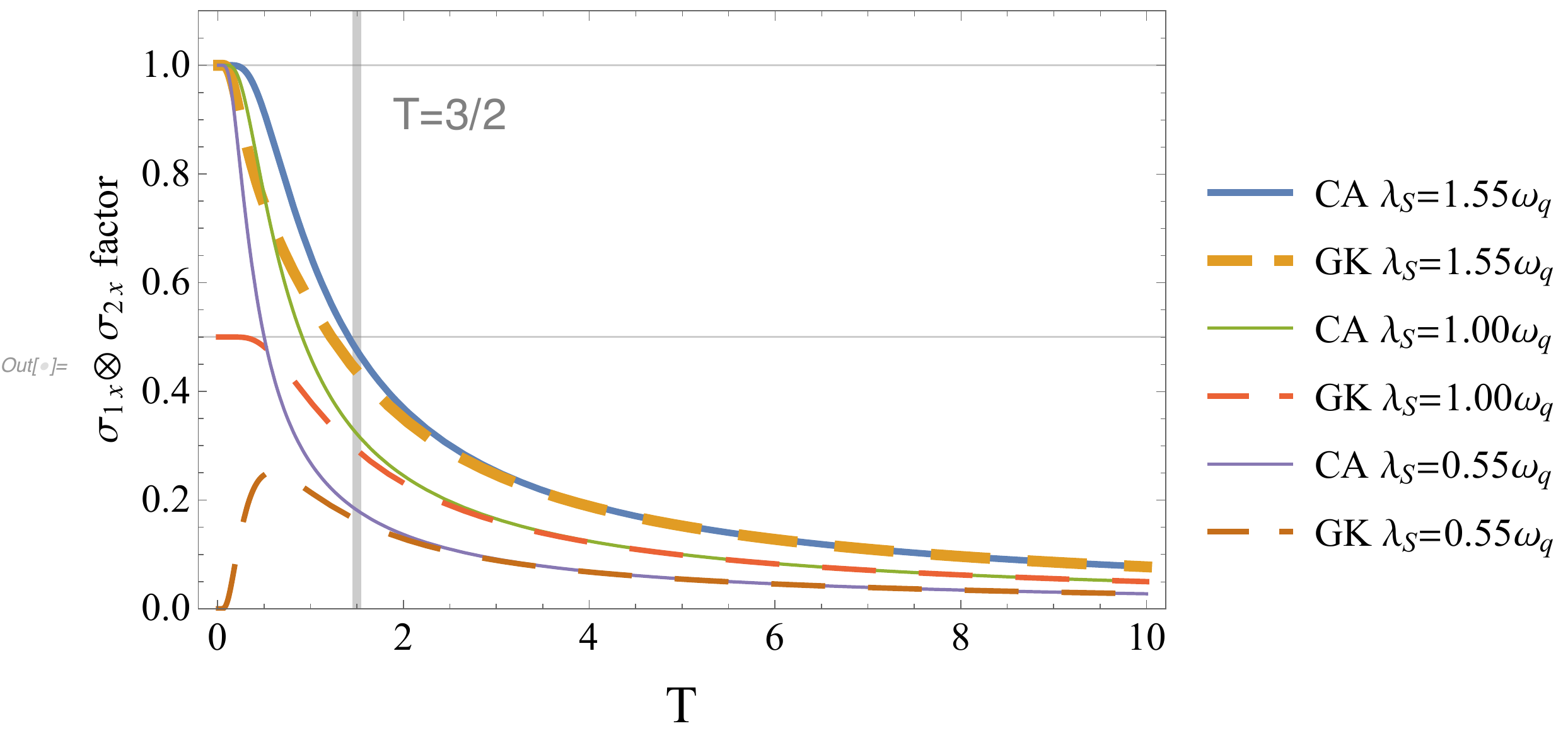}
\caption{\label{fig:Twoqubits} 
$\sigma_{1x} \otimes \sigma_{2x}$ prefactor $\tanh(\tfrac{1}{2}\beta\lambda_S)$ from \eqref{eq:twoqubitMFGS} (CA, solid) and corresponding prefactor $\frac{\sinh\beta\lambda_S}{\cosh\beta\omega_q+\cosh\beta\lambda_S}$ from \eqref{eq:GandKTwoQubits} (GK, dashed) are plotted as functions of  temperature $T=1/\beta$, for three inter-qubit couplings $\lambda_S =1.55\omega_q, 1.00\omega_q$ and $0.55\omega_q$, for $\omega_q=1$. 
While for  large temperatures these two expressions are approximately the same, for low temperatures  they differ very much, especially for smaller inter-qubit couplings $\lambda_S \lesssim \omega_q$. 
For spin-reservoir coupling $\lambda_1 = 4\omega_q = \lambda_2$ and inter-qubit coupling $\lambda_S=1.55$ (thicker lines), Goyal and Kawai (GK) have fully solved the dynamics numerically at inverse temperature value $T=3/2$ (grey) using the method of hierarchical equations of motion (HEOM). Their numerical results indicate that the dynamical steady state $\stst_S (t \to \infty)$ is numerically close to \eqref{eq:OK}. A fully conclusive separation between convergence to either \eqref{eq:GandKTwoQubits} or \eqref{eq:twoqubitMFGS} is not possible at these parameter values, and requires further numerical studies.}
\end{figure}

Consequently we find that
\begin{equation}
	P_{1m}P_{2n}H_SP_{2n}P_{1m}=P_{1m}P_{2n} H^{int}_{12} P_{2n}P_{1m}=\lambda_S\left(P_{1m}\sigma_{1+}P_{1m} \, P_{2n}\sigma_{2-}P_{2n}+\text{H.c.}\right),
\end{equation}
for $m = \pm$ and $n = \pm$. 
With $P_{\alpha \pm}\sigma_{\alpha +}P_{\alpha \pm}=P_{\alpha \pm}\sigma_{\alpha -}P_{\alpha \pm}=\tfrac{1}{2}\sigma_{\alpha x}$ it follows directly that
\begin{equation}
	\tilde{\rho}_S=\exp\left[-\beta\sum_{m, n =\pm}P_{1m}P_{2n} H^{int}_{12}  P_{2n}P_{1m}\right]=\exp\left[- \frac{1 }{ 2} \beta \lambda_S \, \sigma_{1x} \otimes \sigma_{2x}\right].
\end{equation}
Normalization then yields the strong coupling \MFstate~for the two coupled qubit system stated in the main text \eqref{eq:twoqubitMFGS_JC}, i.e.
\begin{equation}\label{eq:twoqubitMFGS}
	\rho_S=\frac{1}{4}\left(1- \sigma_{1x} \otimes \sigma_{2x} \tanh(\tfrac{1}{2}\beta\lambda_S)\right),
\end{equation}
which is independent of $\omega_q$.  Again the operator pre-factor $\tanh(\tfrac{1}{2}\beta\lambda_S)$ is plotted as a function of  temperature $T$  in Fig.~\ref{fig:Twoqubits}, for various values of inter-qubit coupling $\lambda_S$.

Once again, this result can be contrasted with the form of the equilibrium state conjectured by Goyal and Kawai \cite{Goyal2019}, generalised to the case of two reservoirs, which takes the form
\begin{equation}\label{eq:GandKTwoQubits}
	\varrho_S^{GK}=\sum_{m=\pm}\sum_{n=\pm}P_{1m}P_{2n}\tau_SP_{2n}P_{1m}=\frac{1}{4}\left(1- \sigma_{1x} \otimes \sigma_{2x} \frac{\sinh\beta\lambda_S}{\cosh\beta\omega_q+\cosh\beta\lambda_S}\right),
\end{equation}
which is clearly dependent on $\omega_q$ giving it additional structure in comparison to  \eqref{eq:twoqubitMFGS} which is independent of $\omega_q$. Both \eqref{eq:GandKTwoQubits} and the result derived here, \eqref{eq:twoqubitMFGS}, are shown in Fig.~\ref{fig:Twoqubits} where \eqref{eq:GandKTwoQubits}'s dependence on $\omega_q$ leads to peak when $\lambda_S \ll \omega_q$, which is not observed for \eqref{eq:twoqubitMFGS}. 

A comparison can be attempted between the above two conjectured/derived states, and the numerical results of Goyal and Kawai for the steady state  $\lim_{\lambda\to\infty}\stst_S(t\to\infty)$. They obtain the latter  by use of  hierarchical equations of motion (HEOM) methods to numerically solve the master equation for the two qubit system, and in the large time limit obtain the steady state of the system in the limit of strong coupling between the qubits and their respective reservoirs. 
For their chosen parameter values \cite{Goyal2019}, qubit energy $\omega_q=1$, qubit-bath couplings $\lambda_1 = 4\omega_q=\lambda_2$, inter-qubit coupling $\lambda_S=1.55\omega_q$,  and  temperature $T=3/2$, the comparison between the master equation steady state and the conjectured form \eqref{eq:GandKTwoQubits} is measured by the fidelity, $\mathcal{F}(\rho,\rho')=\tr\left[\sqrt{\sqrt{\rho}\rho'\sqrt{\rho}}\right]^2$, and Goyal and Kawai found very close agreement. 

However, for the parameter values chosen, the states $\varrho_S^{GK}$ and $\rho_S$  from \eqref{eq:twoqubitMFGS_JC} are also approximately equal (fidelity $>0.9995$), and hence the numerical convergence of the steady state $\lim_{\lambda \to \infty} \stst_S (t \to \infty)$ to the \MFstate~$\rho_S$ is equally plausible, see Fig.~\ref{fig:Twoqubits}. 
Clearer disambiguation of convergence to either $\varrho_S^{GK}$ or $\rho_S$ could be achieved at lower temperatures~\footnote{R. Kawai, private communication 2019/2020}, however, the HEOM method has its own convergence restrictions that may limit the range of parameters that can be explored. 

\end{document}